\documentclass[prb,preprint,aps,showpacs]{revtex4} 
\usepackage{graphicx} 
\usepackage{tabularx}
\usepackage{dcolumn} 
\usepackage{color}
\usepackage{bm}
\usepackage{amsmath}
\usepackage{amssymb}

\begin{document} 
 \title{
Novel $H_{\rm c2}$ suppression mechanism in  
a spin triplet superconductor \\
-- Application to UTe$_2$--}
\author{Kazushige Machida} 
\affiliation{Department of Physics, Ritsumeikan University, Kusatsu 525-8577, Japan} 

\date{\today} 

\begin{abstract}
A novel $H_{\rm c2}$ suppression mechanism is theoretically proposed in a spin triplet superconductor (SC)
with equal spin pairs. We show that the upper critical field $H_{\rm c2}$ can be reduced from the orbital depairing limit $H^{\rm orb}_{\rm c2}$
to arbitrarily small value, keeping the second order phase transition nature. This mechanism is sharply
different from the known Pauli-Clogston limit for a spin singlet SC where the reduction is limited to $\sim$0.3$H^{\rm orb}_{\rm c2}$
with the first order transition when the Maki parameter goes infinity. This novel $H_{\rm c2}$ suppression mechanism is applied  to UTe$_2$,
which is a prime candidate for a spin triplet SC, to successfully analyze the $H_{\rm c2}$ data for various crystalline orientations both under ambient and applied pressure, and
to identify the pairing symmetry. It is concluded that the non-unitary spin triplet state with equal spin pairs is realized in UTe$_2$, namely
$(\hat b+i\hat c)k_a$ in $^3$B$_{\rm 3u}$ which is classified under finite spin orbit coupling scheme.
\end{abstract}

\maketitle

\section{Introduction}
It is well known that the upper critical field $H_{\rm c2}$ is suppressed by the so-called Pauli-Clogston mechanism
for spin-singlet superconductors through the Zeeman effect, which is characterized by the Maki parameter $\alpha_{\rm Maki}$~\cite{sant-james}.
For larger $\alpha_{\rm Maki}\geq 1$ the phase transition at $H_{\rm c2}$ becomes first order from usual second order phase transition,
and the reduction of $H_{\rm c2}$ is saturated, tending gradually to a lower bound $\sim$$0.3H^{\rm orb}_{\rm c2}$ with  the orbitally limited 
$H^{\rm orb}_{c2}$
toward $\alpha_{\rm Maki}$$\rightarrow$$\infty$~\cite{214}.
In contrast, there exists no known $H_{\rm c2}$ suppression mechanism for a spin-triplet superconductor, except that the
d-vector is firmly locked parallel to the external field direction. This situation is the same as in the spin-singlet case.
Thus it would be quite surprising if we see that $H_{\rm c2}$ is suppressed
by an external field when the d-vector is rotated perpendicular to it. 

Recently, much attention has been focused on a newly found heavy Fermion superconductor 
UTe$_2$~\cite{ran,aoki0,review,review1}.
Since the upper critical field far exceeds the Pauli paramagnetic limit set by
$H_{\rm p}=1.75T_{\rm c}\sim 3.5$T for all crystalline directions,
it is expected that the realized pairing symmetry belongs to a spin-triplet category~\cite{review,review1}.
However, details of the pairing function  remain unidentified and are much debated until now~\cite{review,review1}. 
Because of the rich internal degrees of freedom in the spin-triplet pairing function which consists of the
spin $SO(3)^{\rm spin}$ and orbital $D_{\rm 2h}^{\rm orbital}$ parts in general, 
the multiple superconducting states are expected to exist.

Indeed recent several experiments including specific heat~\cite{rosuel}  and flux flow measurements~\cite{sakai} unambiguously
demonstrate that at least three phases exist in the $H$-$T$ plane ($H\parallel$$b$) at the ambient pressure,
in addition to previously known multiple phase diagrams in the $H$-$T$ plane under pressure $P$~\cite{pressure0,pressure1,pressure2,pressure3,pressure4}.
These observed multiple phase diagrams are a hallmark of a spin-triplet superconductor (SC)
and similar to UPt$_3$~\cite{taillefer,sauls,upt3,ohmi,yo,tsutsumi1} another spin-triplet SC with the three phases; A, B and C in  the $H$-$T$ plane
and also the superfluid $^3$He which consists of the A and B phases in the $P$-$T$ plane~\cite{3he,mizushima}.

It is instructive to remind of the fact that in the A phase in the superfluid $^3$He
the transition temperature $T_{\rm c}$ splits into two; the A$_1$ phase with $T_{\rm c1}$ and A$_2$ phase with $T_{\rm c2}$
under applied field $H$. The former (latter) shits up (down) linearly in $H$
up to at least 16T~\cite{ishimoto} because  the spin $\uparrow\uparrow$ ($\downarrow\downarrow$) pairs
gain (loose) the magnetic energy.

Here since we are advocating that in UTe$_2$ the A$_1$ and A$_2$ like non-unitary pairing
state is able to describe a variety of exotic phenomena, including the 
$T_{\rm c}$ increase with increasing $H(\parallel$$b)$~\cite{review,review1}.
This particular phenomenon is akin to the $T_{\rm c1}(H)$ rise of the A$_1$ phase
under $H$ mentioned. Then it is natural to ask where  the decreasing $T_{\rm c}$ for the A$_2$ phase 
with $\downarrow\downarrow$ pairs
exists in the $H$-$T$ phase diagram because the A$_1$ and A$_2$ are originated from the same
mother A phase. 

We are motivated by the recent intriguing two experimental papers~\cite{hc2',pressure}:
The first paper~\cite{hc2'} reports the orientational dependences of the initial slopes $dH_{\rm c2}/dT$ at $T_{\rm c}$
and $H_{c2}$ for all crystalline angles as will be shown later (see Fig.~\ref{hc2}).
Since according to a standard formula: 
$H^{\rm WHH}_{\rm c2}(T\rightarrow 0)\sim-0.7({dH_{\rm c2}/ dT})_{T_{\rm c}}\cdot T_{\rm c}$,
given by Wertharmer, Helfand, and Hohenberg~\cite{whh}, $H_{\rm c2}$ must be proportional to the initial slope.
While along the $c$-axis $H^c_{\rm c2}=17$T nearly coincides with $-({dH_{\rm c2}/dT})_{T_{\rm c}}=6$T/K by 
multiplying a factor 3 with $T_{\rm c}=2.1$K, the other directions 
$H^a_{\rm c2}=12$T and $H^b_{\rm c2}=23$T should be compared with 
$H'^a_{\rm c2}=-15$T/K and $H'^b_{\rm c2}=-23$T/K. Thus the actual $H^a_{\rm c2}$ and
$H^b_{\rm c2}$ are far below the expected 
$H^a_{\rm c2}\sim 45$T and $H^b_{\rm c2}\sim 75$T by multiplying the same factor 3.
This implies some unknown mechanism to exist in order to explain these large $H_{\rm c2}$ suppressions
which should be field-orientation dependent.

The other paper~\cite{pressure} reports the impressive pressure evolution of 
the $H$-$T$ multiple phase diagrams for $H$$ \parallel$$b$:
The high field phase SC2 in their terminology above $H=14$T in the ambient pressure progressively goes down toward lower 
field and eventually reaches the $H=0$ line and is stabilized at higher $T$ than the lower field phase SC1
at around $P=0.19$GPa. Together with other pressure experiments~\cite{pressure3,pressure1,pressure4,pressure2} 
this pressure evolution of the multiple phase diagrams is seemingly independent
of the above $H_{\rm c2}$ suppression phenomenon,
but in this paper we show a deep internal interdependence between them 
due to the inherent nature of the pairing symmetry realized in
UTe$_2$. These analyses lead us to believe in identifying our pairing symmetry.

\begin{figure}
\begin{center}
\includegraphics[width=12cm]{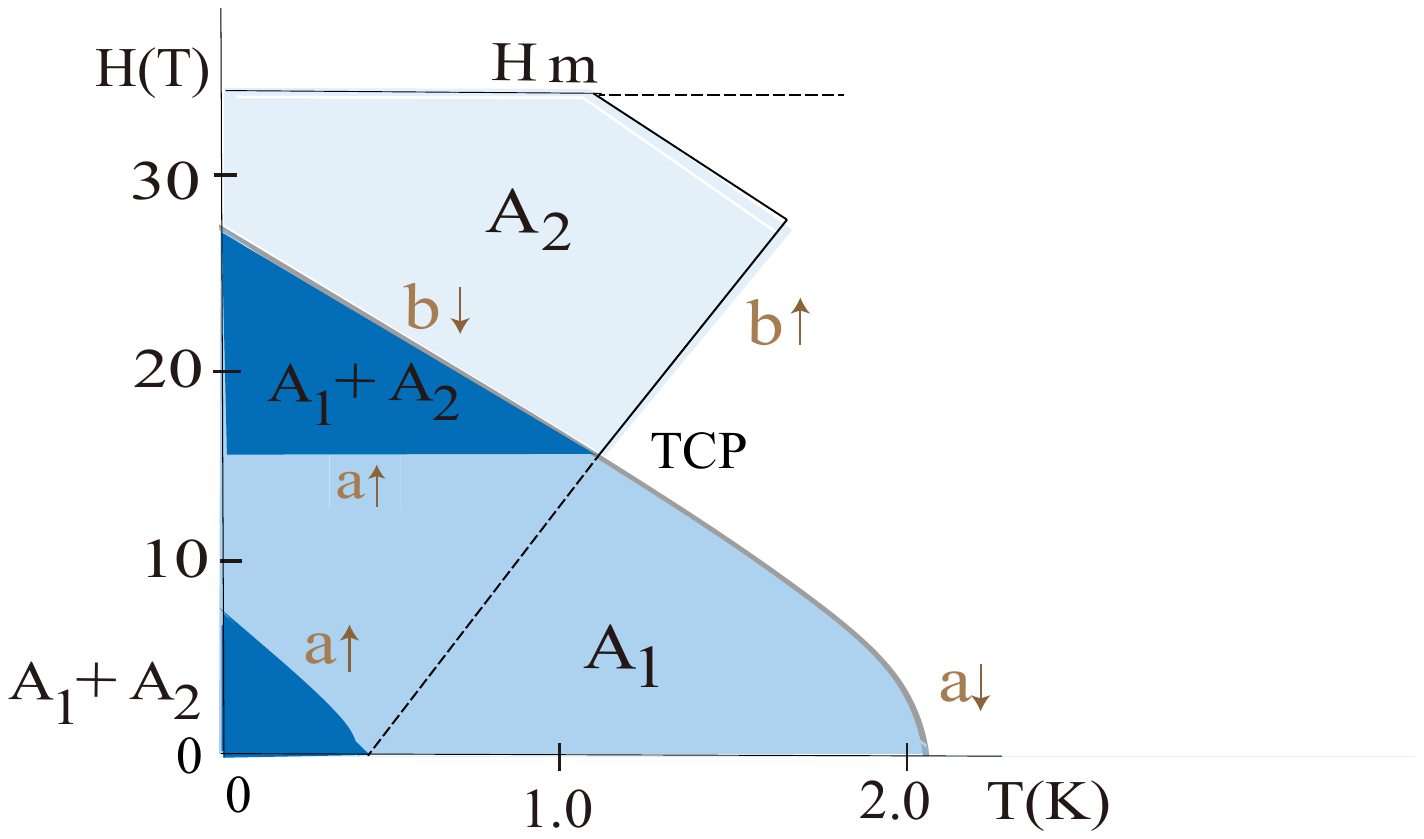}
\end{center}
\caption{\label{schematic}
Schematic $H$-$T$ phase digram for $H\parallel b$-axis~\cite{machida5,machida6}.
In the A$_1$(A$_2$) phase the spin polarization $\bold S$ points to the antiparallel (parallel) direction 
along the $a$-axis at low fields and turns to the parallel (antiparallel) to the $b$-axis in higher fields
above the d-vector rotation field $H_{\rm rot}$ denoted by TCP. $H_{\rm m}$ is the first order metamagnetic transition.
The dotted line inside the A$_1$ phase is the hypothetical transition line for the A$_2$ phase.
}
\end{figure}

Since the present paper belongs to a series of our papers on UTe$_2$~\cite{machida1,machida2,machida3,machida4,machida5,machida6,tsutsumi},
it might be useful to summarize the main points achieved so far
and to explain the background for investigations of the novel $H_{\rm c2}$ suppression mechanism.
It will be turned out, however, that this mechanism is applicable to a spin-triplet superconductor
characterized by an equal spin state in general. 

As shown schematically in Fig.~\ref{schematic} under the ambient pressure~\cite{machida5,machida6} the
phase diagram in the $H$-$T$ plane consists of the two phases 
A$_1$  and A$_2$, corresponding to low field phase SC1 and high field phase SC2 respectively. 
The A$_1$ (A$_2$) phase is described by the Cooper pair spin $\downarrow\downarrow$
($\uparrow\uparrow$) whose spin-quantization axis is anti-parallel (parallel) to the magnetic easy axis $a$
at lower fields although we do not know the exact origin of this T$_{\rm c}$ splitting mechanism at $H=0$. This is consistent with the Knight shift (KS) experiment~\cite{matsumura};
The KS below $T_{\rm c}$ decreases for $H\parallel a$-axis because the $\downarrow\downarrow$ pairs diamagnetically respond
to applied field, meaning that these $\downarrow\downarrow$ pairs are energetically unfavorable under $H$.

In the higher fields above $H>H_{\rm TCP}=14$T, 
the A$_2$ reappears with the spin quantization axis
along the $b$-axis due to the d-vector rotation~\cite{ishida1,ishida2,ishida3,ishida4,ishida5,kinjo,kitagawa}.  
The four second order phase transition lines meet at $H_{\rm TCP}$, constituting the
tetra-critical point (TCP) above which $H^b_{\rm c2}$ becomes having a positive sloped $H_{\rm c2}$,
leading to the strong $H_{\rm c2}$ enhancement.
This is caused by the Cooper pair polarization $\bold S$ becomes pointing to the positive direction relative
to the $b$-axis magnetization $M_b(H)$ which is parallel to the external field $H$$\parallel$$b$-axis
in order to gain the magnetic energy arising the coupling between the Cooper pair polarization and 
magnetization.
Here $H_{\rm TCP}$ corresponds to the field $H_{\rm rot}$ that the d-vector rotation is completed~\cite{ishida1,ishida2,ishida3,ishida4,ishida5,kinjo,kitagawa}.
This understanding is consistent with KS experiment where the KS drop below $T_{\rm c}(H)$ gradually
ceases and remain unchanged as $H$ grows above $H_{\rm rot}$=14T~\cite{ishida5}. 
In our papers the construction of the phase diagram, including  the strong $H_{\rm c2}$ enhancement,
 is explained in detail.

Basically it is due to the fact that under an applied field, $T_{\rm c}(H)\propto {\bold M}_b(H)\cdot \bold S$ in $H>H_{\rm rot}$ 
through the generic coupling between the Cooper pair polarization $\bold S$ and the field-induced magnetization vector ${\bold M}(H)$.
Namely this is deeply rooted to the inherent nature of the non-unitary  pairing symmetry with the equal spin pairs.
This strong $H_{\rm c2}$ enhancement phenomenon is analogous to the $T_{\rm c}(H)$ increase
of the superfluid $^3$He-A phase under $H$ as mentioned above.
There is no corresponding $H_{\rm c2}$ suppression phenomenon identified so far in UTe$_2$.
In other words, $T_{\rm c}(H)$ strongly decreases as $H$ increases.
This phenomenology is highly expected to occur in UTe$_2$ once we assign the  A$_1$ and A$_2$-like
phases analogous to the superfluid $^3$He-A phase because $T_{\rm c}(H)$ increase and decrease
occur in pair and are tightly connected.
If found in UTe$_2$, it strengthens our scenario based on the non-unitary pairing state
and gives an important clue to finally pin down the pairing symmetry realized in UTe$_2$.
We warn here that the $T_{\rm c}$-splitting at $H=0$ and the $H_{\rm c2}$ suppression are different phenomena.
The former is related to the pairing mechanism while the latter occurs only under the external field.
Thus in this paper we are not going into details on the origin of the $T_{\rm c}$-splitting, and just assume
that the A$_1$ phase is characterized by the spin $\downarrow\downarrow$ pairs.

The arrangement of the paper is as follows:
We first explain the  $H_{\rm c2}$ suppression in Sec. I based on a
Ginzburg-Landau (GL) formalism. This section is quite generic valid for the
spin $\downarrow\downarrow$ pair state.
We start to analyzing the experimental data to prove that
this novel suppression mechanism is in fact working in UTe$_2$ in Sec. III.
 Then we go on to examine the multiple phase diagrams under pressure in Sec. IV
 and to see that this suppression mechanism also works together with the previously identified 
 $H_{\rm c2}$ enhancement  mechanism. This lets us
 better understand the pressure evolution of these multiple phase diagrams
 and assures us the present non-unitary pairing symmetry realized in this material.
 We further study these points in Sec. V.
 In Sec. VI, discussions  are given from more general point of view 
 and in the final section we devote to conclusion and summary.

\subsection{Nomenclature of A$_1$, A$_2$, and A$_0$}

Before embarking on the detailed studies, we clarify the nomenclature used in the present paper:
The notations which denote three superconducting phases and its mixtures are
borrowed from the superfluid $^3$He-A phase~\cite{3he,mizushima,ishimoto}.
In fact, as explained in a series of papers~\cite{machida1,machida2,machida3,machida4,machida5,machida6,tsutsumi}
this analogy is quite appropriate and useful, but we need to understand several important differences
in the fundamental aspects.
Since we assume a spin triplet pairing, there exist three kinds of phases, spin up $\Delta_{\uparrow}$,
spin down $\Delta_{\downarrow}$, and spin zero $\Delta_{0}$ phases  relative to a spin quantization axis,
corresponding to $S_z=+1, -1, 0$ respectively.
In order to fully characterize the realized states in $H$-$T$-$P$ space we have to specify the
spin component and the associated spin quantization axis.
For example, under an applied field the d-vector may rotate by changing the Cooper pair spin direction
so that the associated spin quantization axis alters correspondingly as shown in Fig.~\ref{schematic}.
We characterize each phase with the spin direction and the associated spin quantization axis 
denoted by the principal crystalline axes $a$, $b$, and $c$. We also note that 
the lower temperature phases below the second transition under a fixed field are
always the mixture of the high T phase and low T phase. For example, 
in Fig.~\ref{schematic} the low T phase denoted as $A_1+A_2$ are the mixture of 
$A_1$ with $a \downarrow$ and $A_2$ with $a \uparrow$  where $a$ is the spin quantization axis
while above TCP $A_2$ with $b \uparrow$ and $A_1$ with $b \downarrow$ are mixed in high fields.
Here the terminology of A$_1$ and A$_2$ is used to merely distinguish 
two kinds of the spin pairs $\uparrow\uparrow$ and $\downarrow\downarrow$
where the spin quantization axis depends on the situation.
In the superfluid $^3$He-A phase, the spin quantization axis is always along the applied magnetic field direction,
a situation quite different from our cases in UTe$_2$.
The orbital part of the pairing function is different: $p_x+ip_y$ type with the point nodes in $^3$He-A phase
while it is not determined in UTe$_2$.

\subsection{Preliminaries to non-unitary triplet pairing} 

We briefly recapitulate  our previous framework in order to facilitate finding 
the novel $H_{\rm c2}$ suppression mechanism and apply it for UTe$_2$.
Starting with the general Ginzburg-Landau (GL) theory 
for a spin triplet state~\cite{machida1,machida2,machida3,machida4,machida5,machida6,tsutsumi},
we make the following assumptions in the present paper:
We assume a nonunitary A-phase-like pairing state described by the complex $\bf d$-vector:
${\bf d}(k)=\phi(k){\boldsymbol \eta}=\phi(k)({\boldsymbol \eta}'+i{\boldsymbol \eta}'')$
with ${\boldsymbol\eta}'$ and ${\boldsymbol \eta}''$ real vectors.
$\phi(k)$ is the orbital part of the pairing function  
which is not specified in the main body because it is irrelevant,
and  the last section discusses its form.
The pairing function is obeyed under the overall symmetry 
$SO(3)^{\rm spin}\times D_{2h}^{\rm orbital}\times U(1)^{\rm guage}$
with the spin, orbital, and gauge symmetry, respectively~\cite{machida0,annett},
assuming the weak spin-orbit coupling scheme (SOC)~\cite{ozaki1,ozaki2}.
This scheme is justified by the experimental fact that 
the d-vector rotation begins from the low fields, $\sim$1 T for the $c$-axis~\cite{ishida3}, and $\sim$5 T
and its gradual rotation is completed at 15T for the $b$-axis~\cite{ishida2}.
This indicates that the spin-orbit coupling strength, which locks the d-vector 
to crystalline lattices, is finite and anisotropic, corresponding to these magnetic field values. 
Thus the $SO(3)^{\rm spin}$ symmetry is weakly broken, which is taken into account perturbationally.
We note that in the strong SOC scheme
the gradual d-vector rotation spanning over 10T is not possible because the d-vector locking energy 
is infinitely strong. 

We assume the observed
ferromagnetic fluctuations in various experimental methods~\cite{ran,sonier,tokunaga1,tokunaga2,furukawa,tokunaga-prl} 
slower than the Cooper pair formation time to stabilize the nonunitary triplet pairing state~\cite{sugiyama,ramires}.
According to the recent NMR experiment on high-quality samples,
Tokunaga et al~\cite{tokunaga-prl} discover extremely slow 
longitudinal magnetic fluctuations on their $T_2$ measurements
in the normal state.
 The SO(3)$^{\rm spin}$ triple spin symmetry for the Cooper pair spin space
permits us to introduce a complex
three-component vectorial order parameter ${\boldsymbol \eta}=(\eta_a,\eta_b,\eta_c)$.
The spin space symmetry is weakly perturbed by the 5f localized moments of the U atoms through the ``effective'' 
spin-orbit coupling felt by the Cooper pairs in the many-body sense because the one-body SOC effects associated
with heavy U atoms are already 
taken into account in forming one-body band structure.

\section{H$_{\rm c2}$ suppression}

In order to understand the general H$_{\rm c2}$ suppression mechanism for an equal spin pairing state with
the spin $\downarrow$$\downarrow$ pairs,
we assume the following situations and restrictions:

\noindent
(1) The Cooper pairs with the spin $\downarrow$$\downarrow$ are assumed to appear at T$_{\rm c}$.
The spin quantization axis is defined along the induced component direction of the magnetic moment  M(H$_{\rm ext}$),
which is induced by the external field H$_{\rm ext}$.
Therefore, the Cooper pair spin direction is anti-parallel to the external field direction.

\noindent
(2) These Cooper pairs with the spin $\downarrow$$\downarrow$ are unfavorable energetically 
under H$_{\rm ext}$ relative to the Cooper $\uparrow$$\uparrow$ pairs.
 The Cooper pairs with the spin $\downarrow$$\downarrow$ respond diamagnetically to H$_{\rm ext}$
 whereas the Cooper $\uparrow$$\uparrow$ pairs respond paramagnetically.
This situation is contrasted with the case in the superfluid A$_1$ phase with the spin $\uparrow$$\uparrow$ pairs
(A$_2$ with the spin $\downarrow$$\downarrow$ pairs) whose T$_{\rm c}$ increases (decreases)
by H$_{\rm ext}$ because the Cooper pair spin is free to align along the H$_{\rm ext}$ direction to save the
magnetic energy. This can be neatly described by the GL free energy in terms of 
$\propto \kappa H_{\rm ext}(\Delta^2_{\uparrow}-\Delta^2_{\downarrow})$
where $\Delta_{\uparrow}$ and $\Delta_{\downarrow}$ are the order parameters.
Here the magnetic response is always paramagnetic and T$_{\rm c}$ increases
through the magnetic coupling term above.

\noindent
(3) The Cooper pair spin is assumed to be tightly locked to the induced magnetic moment, 
that is, the external field direction. T$_{\rm c}$ decreases through the magnetic coupling above by
the amount of $\kappa M(H_{\rm ext})$.

\noindent
(4) We only consider the field induced situations by the external applied field to discuss the H$_{\rm c2}$ suppression,
which is independent of the complicated and subtle situations under zero field and the T$_{\rm c}$ splitting mechanism.

Under these assumptions and restrictions, it is easy to derive the H$_{\rm c2}$ expression for the state $\eta$ with
the spin $\downarrow$$\downarrow$ Cooper pairs through the GL free energy as

\begin{eqnarray}
F=a{_0}(T-T_{\rm c}(H_{\rm ext})|\eta|^2 
+K_{a}|D_a\eta|^2+K_{b}|D_b\eta|^2+K_c|D_c\eta|^2.
\label{GL}
\end{eqnarray}


\noindent
where the transition temperature under fields is shifted to $T_{\rm c}(H_{\rm ext})=T_{\rm c}-\kappa M(H_{\rm ext})$
due to the induced moment via the magnetic coupling ($\kappa>0$).
The variation with respect of $\eta^{\ast}$ results in  

\begin{eqnarray}
a_0(T-T_{\rm c}(H_{\rm ext}))\eta+(K_{a}D_a^2+K_{b}D_b^2+K_cD_c^2)\eta=0.
\label{osci}
\end{eqnarray}

\noindent
The upper critical field $H_{\rm c2}$ is given as the lowest eigenvalue of the linearized 
GL equation or Schr\"odinger type equation of a harmonic oscillator~\cite{tinkham} as,


\begin{eqnarray}
H_{{\rm c2},j}(T)=\alpha^j_0(T_{\rm c}-\kappa M(H_{\rm ext})-T)
\label{hc2}
\end{eqnarray}

\noindent
with $j$=$a,b,c$. We suppress the subscript ``ext'' from now on.
$M(H)$ is the field induced part of magnetization, that is, $M(H=0)=0$.
 We have introduced,

\begin{eqnarray}
\alpha^{a}_0&=&{\Phi_0\over 2\pi \sqrt{K_bK_{c}}}a_0,\qquad
\alpha^{b}_0={\Phi_0\over 2\pi \sqrt{K_cK_{a}}}a_0,\nonumber \\ 
\alpha^{c}_0&=&{\Phi_0\over 2\pi \sqrt{K_aK_{b}}}a_0.
\label{mass}
\end{eqnarray}

\noindent
These coefficients determine the initial slopes of the upper critical fields
for each direction.

Expressing Eq. (\ref{hc2}) in a general form by suppressing the index $j$,
we obtain:

\begin{eqnarray}
H_{{\rm c2}}(T)+\alpha_0\kappa M(H_{\rm c2})=\alpha_0(T_{\rm c}-T).
\label{gl}
\end{eqnarray}


\begin{figure}
\begin{center}
\includegraphics[width=12cm]{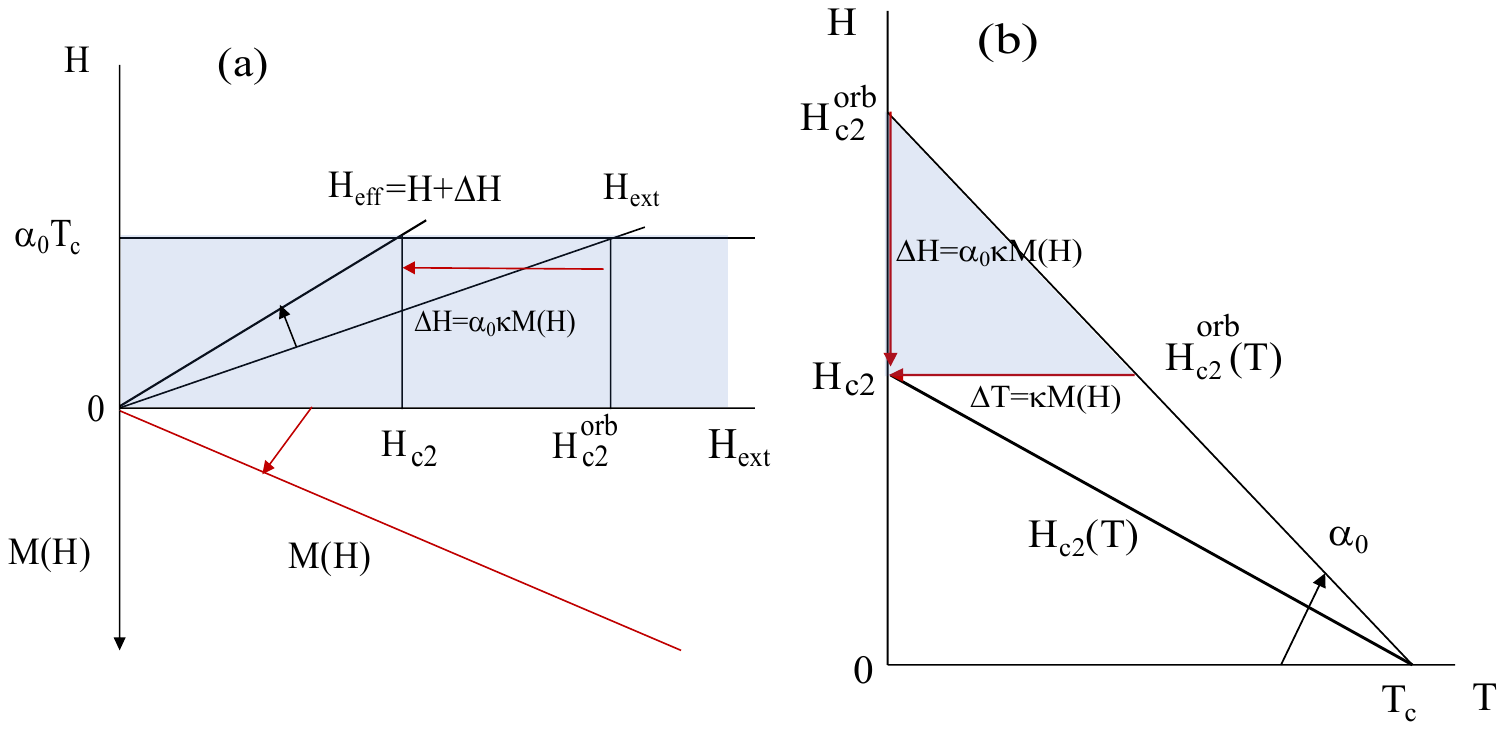}
\end{center}
\caption{\label{gainen}
(a) Schematic figure to explain the $H_{\rm c2}$ suppression at $T=0$. 
$H_{\rm c2}$ is reduced from $H^{\rm orb}_{\rm c2}$ by the amount of 
$H_{\rm eff}=H+\Delta H$ with $\Delta H=\alpha_0\kappa M(H)$.
$M(H)\propto H$ is shown below by the red line. 
(b) $H_{\rm c2}(T)$ is reduced from $H^{\rm orb}_{\rm c2}(T)$
by the amount of $\Delta H$ at $T=0$ and by $\Delta T=\kappa M(H)$ along the $T$-axis.  $\Delta H$, $\Delta T$, 
and $H^{\rm orb}_{\rm c2}(T)$
form a triangle in the $H$-$T$ plane. $\alpha_0$=$|(dH^{\rm orb}_{\rm c2}/dT)_{T_{\rm c}}|$.
}
\end{figure}

\noindent
The right-hand side of Eq.~(\ref{gl}) is now

\begin{eqnarray}
H_{\rm c2}^{\rm orb}(T)=\alpha_0(T_{\rm c}-T)
\label{horb}
\end{eqnarray}

\noindent
for the upper critical field owing to the orbital depairing limit
with $T_{\rm c}$ whose maximum value is given by $H_{\rm c2}^{\rm orb}(T=0)=\alpha_0T_{\rm c}$.
On the left-hand side of Eq.~(\ref{gl}) we define the effective field $H_{\rm eff}$ by

\begin{eqnarray}
H_{\rm eff}(H)=H+\alpha_0\kappa M(H).
\label{heff}
\end{eqnarray}


\noindent
This implies that the effective field $H_{\rm eff}(H)$ increases by 
$\Delta H\equiv\alpha_0\kappa M(H)$ from $H$.

The absolute value of $H_{\rm eff}(T)=H_{\rm c2}(T)+\alpha_0\kappa M(H_{\rm c2})$ is bounded by
$|H_{\rm eff}(T)|\leq H^{\rm orb}_{\rm c2}(T=0)$, that is,
\begin{eqnarray}
|H_{\rm c2}(T)+\alpha_0\kappa M(H_{\rm c2})|\leq H^{\rm orb}_{\rm c2}(T=0)=\alpha_0T_{\rm c}
\label{bound}
\end{eqnarray}


\noindent
for $H_{\rm c2}(T)$ to be a solution of Eq.~(\ref{gl}). The right-hand side is determined
by the material parameters in terms of the Fermi velocity $v_{\rm F} $ through the coherent length $\xi$ and the
transition temperature $T_{\rm c}$.
The upper limit of $H_{\rm c2}(0)$ can be reduced at $T\rightarrow 0$
from $H^{\rm orb}_{\rm c2}(T=0)$, namely,

\begin{eqnarray}
H_{\rm c2}(T)\le H^{\rm orb}_{\rm c2}(T).
\label{aul}
\end{eqnarray}

\noindent
As shown schematically in Fig.~\ref{gainen}(a),
at $T=0$ the orbital limited $H^{\rm orb}_{\rm c2}$ is reduced by
 $\Delta H$ or $\alpha_0\kappa M(H)$ because $H_{\rm eff}$ exceeds the
 allowed region set by $\alpha_0T_{\rm c}$ due to the increment of the effective field.
Figure~\ref{gainen}(b) draws the relation between $H_{\rm c2}$ and $H^{\rm orb}_{\rm c2}$.
It is seen from it that the $T_{\rm c}$ shift corresponds to $\Delta T_{\rm c}=\kappa M(H)$.

It may be convenience for later use to summarize  the enhanced $H_{\rm c2}$ case
for the spin $\uparrow$$\uparrow$ Cooper pair state~\cite{machida5,machida6}.
$H_{\rm eff}=H_{{\rm ext}}-\alpha_0\kappa M(H_{\rm ext})$ in this case, corresponding to Eq.~(\ref{heff}).
$H_{\rm c2}(T)$ is always greater than $H^{\rm orb}_{\rm c2}(T)$ in contrast with Eq.~(\ref{aul}). 
Namely, $H_{\rm c2}(T)\ge H^{\rm orb}_{\rm c2}(T)$.

\subsection{General principle of the $H_{\rm c2}$ suppression}

When the Cooper pair polarization $\bf S$ is parallel to the external field, e.g. the magnetization 
vector ${\bf M}(H)$,
$T_{\rm c}(H)$ increases and consequently the effective field $H_{\rm eff}=H-\alpha_0\kappa M(H)$
is reduced, thus $H_{\rm c2}$ is enhanced over $H^{\rm orb}_{\rm c2}$. 
When $\bf S$ is antiparallel to the external field or ${\bf M}(H)$,
$T_{\rm c}(H)$ decreases and the effective field 
$H_{\rm eff}=H+\alpha_0\kappa M(H)$ increases, thus $H_{\rm c2}$ is suppressed.
The $H_{c2}$ suppression occurs for
 the $\downarrow\downarrow$ pairs.
The $H_{c2}$ suppression and enhancement phenomena indicate the
underlying the Cooper pair state, providing us to a valuable tool to determine the
internal Cooper spin structure together with the well-known Knight shift experiment.
As seen above, the $H_{\rm c2}$ enhancement and reduction occur in pair.
Under applied fields, their $T_{\rm c}(H)$ respond differently,
one is enhanced and the other depressed. 
This is a general principle for an equal spin Cooper pair state, which is analogous to superfluid $^3$He-A phase
as mentioned above.

\section{Analyses of experimental data}

\subsection{$H_{\rm c2}$ vs H$_{c2}/dT$}

Let us analyze the experimental data in UTe$_2$, which motivate the present theory.
It is striking to see the data of the $H_{\rm c2}$ and the initial slopes of $dH_{\rm c2}/dT$
for various crystalline angles $\theta$, and $\phi$ (measured from the $c$-axis to the $a$-axis 
and from the $a$-axis to the $b$-axis, respectively)  shown in Fig.~\ref{hc2}
because $H_{\rm c2}$ is expected to be proportional to
the initial slope. For example, according to Werthamer et al~\cite{whh}:
$H_{c2}=0.7|(dH_{\rm c2}/dT)_{T_{\rm c}}|T_{\rm c}$. This general rule is largely violated in UTe$_2$.
In Fig.~\ref{hc2} the angle dependent $H_{\rm c2}(\theta, \phi)$ and $|(dH_{\rm c2}(\theta, \phi)/dT)_{T_{\rm c}}|$ 
are displayed where the latter multiplied by a factor 1.8 with $T_{\rm c}$=2.1K for the overall consistency. 
We have defined $H^{\rm orb}_{\rm c2}$
for the latter quantity, which characterizes the orbital depairing coming from the Fermi velocity anisotropy.

It is seen that \\
\noindent
(1) Almost all portions are dominated by the regions with $H^{\rm orb}_{\rm c2}$$>$$H_{c2}$,
meaning that the actual $H_{c2}$ is largely suppressed from that expected by $H^{\rm orb}_{\rm c2}$.
Near the $a$-axis the ratio of $H_{\rm c2}/H^{\rm orb}_{\rm c2}$ is smallest $\sim0.3$.\\
\noindent
(2) The two curves of $H^{\rm orb}_{\rm c2}$ and $H_{c2}$ are quite in parallel from the $a$-axis to the $b$-axis, implying that $H_{\rm c2}/H^{\rm orb}_{\rm c2}$
is independent of the angle $\phi$.\\
\noindent
(3) In contrast, $H_{\rm c2}$ is enhanced near the $c$-axis.  The large $H_{\rm c2}$ occurring near the $b$-axis is not discussed
in the present paper (see Refs.~[\onlinecite{machida5}] and [\onlinecite{machida6}]).\\
\noindent
(4) It is possible that $H_{\rm c2}$ is enhanced and becomes arbitrarily large by introducing dirtiness in a system
because the effective coherence length $\xi$ which is proportional to the shorter mean free path $l$, i.e., $\xi\propto l$
which determines the vortex core size and limits $H_{\rm c2}(T\rightarrow0)$~\cite{sant-james}.
However, this is not the case since UTe$_2$ is an unconventional superconductor which is vulnerable for impurity scatterings of various kinds.

\begin{figure}
\begin{center}
\includegraphics[width=12cm]{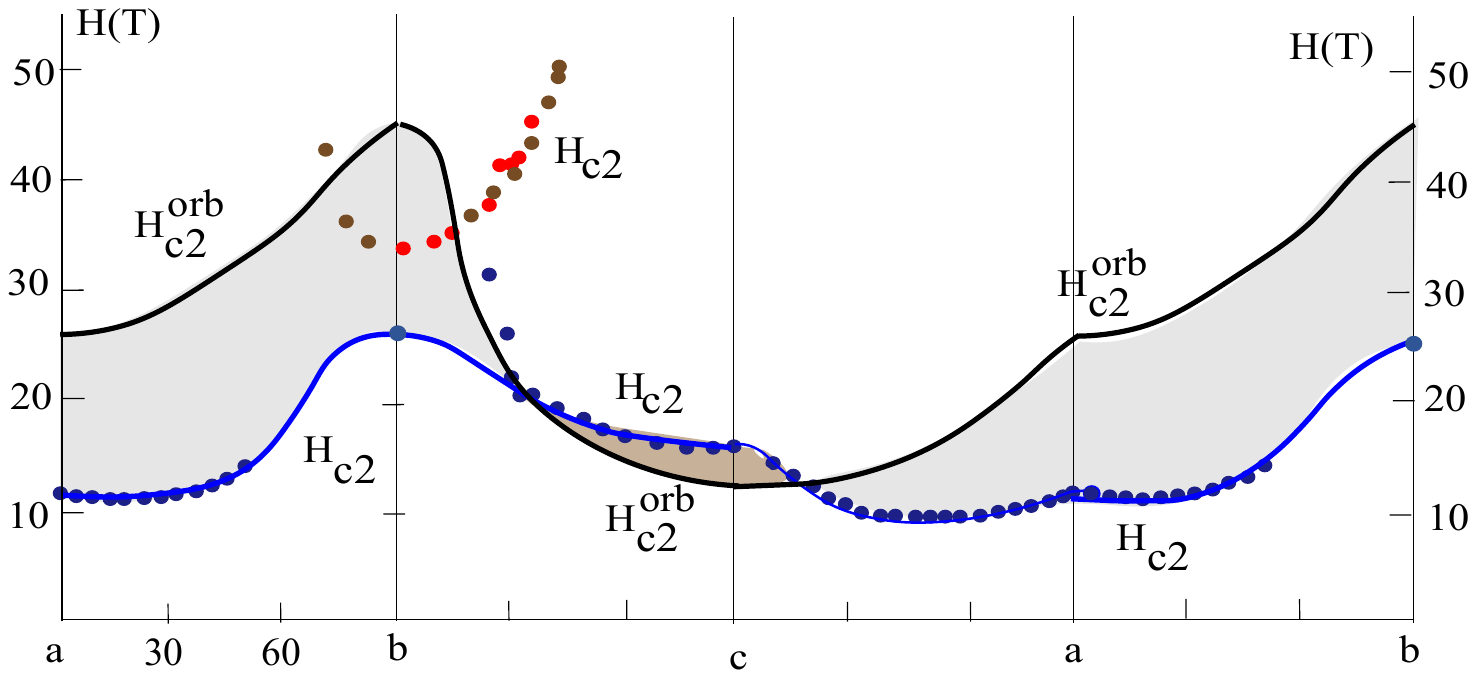}
\end{center}
\caption{\label{hc2}
Comparison of $H_{\rm c2}$ with $H^{\rm orb}_{\rm c2}$ for all field orientations from $a$-axis$\rightarrow$$b$$\rightarrow$$c$$\rightarrow$$a$$\rightarrow$$b$-axis (left to right).
$H^{\rm orb}_{\rm c2}$ is estimated from the initial slopes by $H^{\rm orb}_{\rm c2}=1.8|(dH_{\rm c2}/dT)_{T_{\rm c}}|$.
The color regions indicate the differences between them. The gray (brown) areas show the regions for $H_{\rm c2}<H^{\rm orb}_{\rm c2}$
($H_{\rm c2}>H^{\rm orb}_{\rm c2}$).
The data (dot points) are taken from Aoki et al~\cite{hc2'}. $H_{\rm c2}$ for $H\parallel b$ comes from Refs.~[\onlinecite{rosuel}] and [\onlinecite{sakai}].
 }
\end{figure}

\begin{figure}
\begin{center}
\includegraphics[width=12cm]{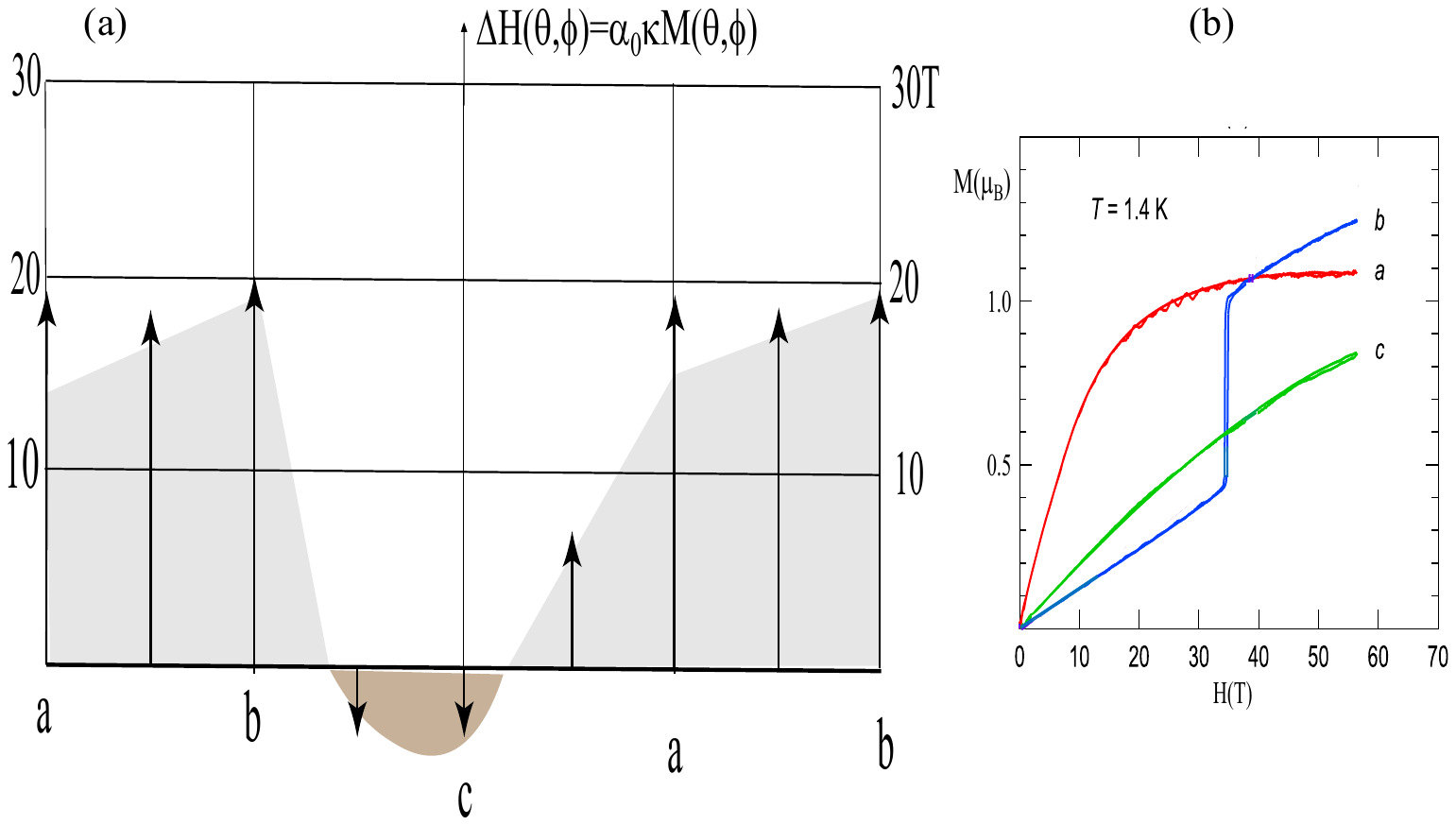}
\end{center}
\caption{\label{deltah}
(a) The difference between $H_{\rm c2}$ and $H^{\rm orb}_{\rm c2}$ shown in Fig.~\ref{hc2} by the gray and brown regions
is compared with the theoretical calculations of  $\Delta H=\alpha_0\kappa M(H)$ for various angles indicated by arrows.
The up arrows (down arrows) show the suppressed (enhanced) $H_{\rm c2}$.
$\alpha_0=|(dH_{\rm c2}/dT)_{T_{\rm c}}|$, $\kappa=2.7$K/$\mu_{\rm B}$, and $M(H)$ from (b).
(b) The magnetization curves $M(H)$ for three principal axes taken from Miyake, et al~\cite{miyake,miyake2}.
We ignore the renormalization of $\alpha_0$ for simplicity and clarity.
}
\end{figure}

Let us estimate the $H_{\rm c2}$ suppression shown in Fig.~\ref{hc2}.
According to Eq.~(\ref{heff}), the reduction $\Delta H=\alpha_0\kappa M(H)$.
We maintain the same value for $\kappa=2.7\mu_{\rm B}$/K as before~\cite{machida5,machida6}.
The initial slope $\alpha_0(\theta,\phi)$ is known from Fig.~\ref{hc2}.
$M(H_{\rm c2})$ can be estimated from the magnetization measurement data~\cite{miyake,miyake2} as shown in Fig.~\ref{deltah}(b).
Then, it is easy to obtain $\Delta H(\theta,\phi)=\alpha_0(\theta,\phi)\kappa M(H_{\rm c2})$ indicated by the arrows in 
Fig.~\ref{deltah}(a). The up (down) arrows corresponds to the $H_{\rm c2}$ suppression (enhancement),
indicating a reasonable agreement.
The values at the three principal axes $a$, $b$, and $c$ are derived later in more details.

We point out a fact that the $H_{\rm c2}$ suppression from the $a$-axis to the $b$-axis
is relatively constant and explain as follows:
The suppression $\Delta H=\alpha_0\kappa M(H)$ consists of $\alpha_0(\phi)$ and $M(\phi)$.
$\alpha_0(\phi)$ is given by the effective mass model:  $$\alpha_0(\phi)=1/\sqrt{m_a\cos^2(\phi)+m_b\sin^2(\phi)},$$
while the angle dependence of the magnetization is described by the so-called elliptic formula~\cite{machida4},
$$M(\phi)=\sqrt{M_a\cos^2(\phi)+M_b\sin^2(\phi)}.$$
Thus if at $\phi=0$ and $90^{\circ}$, $\Delta H(\phi)$ coincides each other, $\Delta H(\phi)$ becomes angle-independent, which is
approximately obeyed by the experimental data.

\subsection{H$_{\rm c2}$ for three principal axes}

\subsubsection{H$^a_{c2}$ in  $H\parallel a$}

As shown in Fig.~\ref{hc2abc}(a), $H^{\rm orb}_{\rm c2}$ $\parallel a$-axis tends approximately to 30T,
but the actual $H^a_{\rm c2}$$\sim$8T. Note that above 8T, $H^a_{\rm c2}$ is slightly enhanced by the metamagnetic transition,
we do not discuss it here (see Tokiwa et al~\cite{tokiwa} and also Shimizu et al ~\cite{shimizu4} for details).
Therefore, the $H_{\rm c2}$ reduction amounts to 8T/30T=0.28. 
At $H$=8T, the reduction of $\Delta T=\kappa M_a(8T)=2.7(K/\mu_{\rm B})\times 0.6\mu_{\rm B}=1.6$K
by reading off from Fig.~\ref{deltah}(b),
and $\Delta H=\alpha^a_0 \Delta T=15(T/K)\times 1.6$K=24T. This leads to 
$H_{\rm c2}=H^{\rm orb}_{\rm c2}-\Delta H=30T-24T=6T$, roughly coinciding with our estimate
$H_{\rm c2}\sim$8T.
In Fig.~\ref{hc2abc}(a) the red triangle indicates this reduction process.
According to the general principle of the $H_{\rm c2}$ suppression mentioned above,
this reduction occurs only for the Cooper pair polarization opposes to the field direction, namely the $a$-axis.
In our assignment $\bf S$ is antiparallel to the $a$-axis which is indeed
consistent with the recent Knight shift experiment by Matsumura et al~\cite{matsumura} (see Ref.~[\onlinecite{machida5}] on this point
for detailed discussion).

\subsubsection{H$^b_{c2}$ in  $H\parallel b$}

\begin{figure}
\begin{center}
\includegraphics[width=14cm]{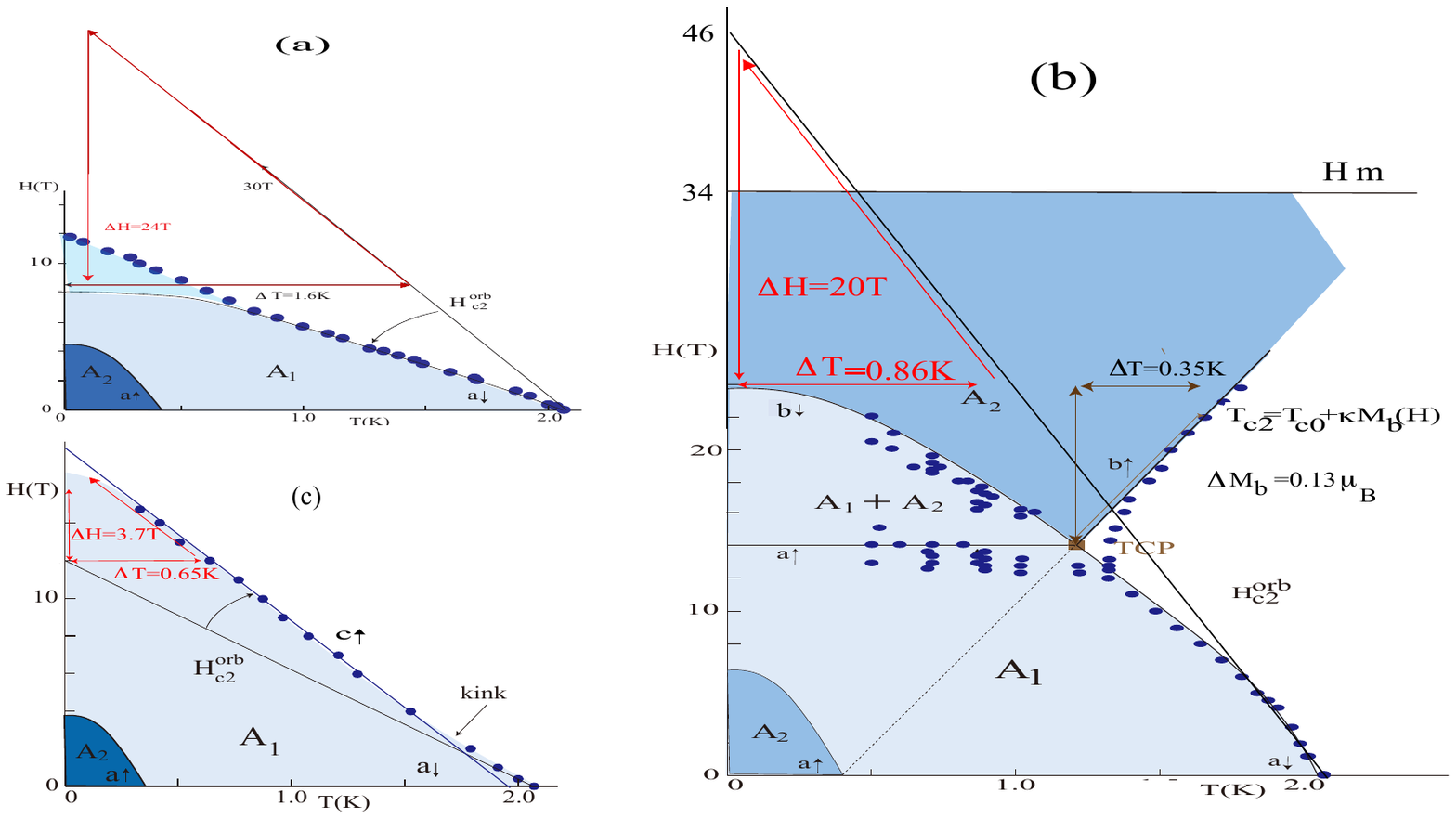}
\end{center}
\caption{\label{hc2abc}
(a) $H^a_{\rm c2}(T)$ for the $a$-axis:  $H^{\rm orb}_{\rm c2}(T\rightarrow0)$ tends to $\sim$30T.
The red triangle shows the $H_{\rm c2}$ reduction. The data are taken from Tokiwa et al~\cite{tokiwa}.
Note that a slight enhancement of $H^a_{\rm c2}(T)$ 
in the high field region is due to
the metamagnetic transition along the $a$-axis above 8T~\cite{tokiwa,shimizu4}.
(b) $H^b_{\rm c2}(T)$ for the $b$-axis:  The red triangle shows the $H_{\rm c2}$ reduction. TCP at 14T denotes the 
teta-critical point
where the four second order transitions meet, corresponding to the d-vector rotation point.
The spin polarization $\bf S$ antiparallel to the $a$-axis at low $H$. The A$_1$ phase changes into the state with $\bf S$ being
antiparallel to the $b$-axis above TCP. The positive sloped $H^b_{\rm c2}(T)$ in the A$_2$ phase above 14T with 
$\bf S$ parallel to the $b$-axis is enhanced with the rate denoted by the triangle with brown color there. 
$H_{\rm m}$ shows the
meta-magnetic transition where A$_2$ terminates. The data points come from the experiments~\cite{sakai}.
(c) $H^c_{\rm c2}(T)$ for the $c$-axis: The red triangle shows the $H_{\rm c2}$ enhancement.
Above 4T denoted by kink, $\bf S$ changes from $a$-antiparallel to $c$-parallel.
$H^{\rm orb}_{\rm c2}(T\rightarrow0)\sim$12T is enhanced.
The data are taken from Tokiwa et al~\cite{tokiwa}.
}
\end{figure}

We continue the same analysis for $H$$\parallel$$b$-axis.
As shown in Fig.~\ref{hc2abc}(b), $H^{\rm orb}_{\rm c2}$ tends to 46T.
However, the actual $H_{\rm c2}$$\sim$24T for the low field phase denoted by the A$_1$ phase~\cite{rosuel,sakai}.
At $H$=24T, $M_b(H=24T)=0.32\mu_{\rm B}$ read from Fig.~\ref{deltah}(b), leads to 
$\Delta T=\kappa M_a(24T)=2.7(K/\mu_{\rm B})\times 0.32\mu_{\rm B}=0.86$K.
Then the resulting $\Delta H=\alpha^b_0 \Delta T=23(T/K)\times 0.86$K=20T.
Thus $H_{\rm c2}=H^{\rm orb}_{\rm c2}-\Delta H=46T-20T=26T$, roughly coinciding with our estimate
$H_{\rm c2}\sim$24T.
Here it is important to understand that along $H^b_{\rm c2}$ line above the tetra-critical point (TCP)
in Fig.~\ref{hc2abc}(b) corresponding to the d-vector rotation point~\cite{ishida4,ishida5},
 $\bf S$ points to the antiparallel direction 
to the field orientation denoted as $b$$\downarrow$ in Fig.~\ref{hc2abc}(b). Therefore, $H_{\rm c2}$ is suppressed.
This is contrasted with the positive sloped $H_{\rm c2}$ above TCP.
After the d-vector rotation, the spin polarization $\bf S$ in this high field phase A$_2$
becomes parallel to the field direction denoted as $b$$\uparrow$ there, which is consistent with the KS experiments~\cite{ishida3,ishida4,ishida5}. 
Thus the magnetization works to
enhance $H_{\rm c2}$. We can estimate its slope as follows:
With increasing field from TCP at $H$=14T to, say 24T,
the $T_{\rm c}$ shift $\Delta T=\kappa \Delta M_b=2.7(K/\mu_{\rm B})\times 0.13\mu_{\rm B}=0.35$K
with the magnetization change $\Delta M_b=0.13\mu_{\rm B}$.
This give rise to a correct slope as shown in Fig.~\ref{hc2abc}(b).
Above TCP, $H_{\rm c2}$ splits into the two $H_{\rm c2}$ curves, one is depressed and the other enhanced,
a situation similar to the $T_{\rm c}$ splitting in the superfluid $^3$He A phase.
Indeed two systems UTe$_2$ and the superfluid $^3$He A phase under  applied field are quite analogous
in this respect. This analogy is an important clue to fully understand the physics in UTe$_2$.

\subsubsection{H$^c_{\rm c2}$ in $H\parallel c$}

Even though at $H$=0 the A$_1$ phase is characterized by the $\downarrow\downarrow$
parallel to the $a$-axis, the applied field tends to the spin polarization $\bf S$ toward the
$c$-axis by rotating the d-vector. This is verified by the Knight shift experiments~\cite{ishida3,ishida4} where
KS becomes to the normal state value below $T_{\rm c}$ and $\bf S$ turns parallel to
the $c$-axis around $H$=3T. This is precisely where $H^c_{\rm c2}$ exhibits a kink,
above which it exceeds $H^{\rm orb}_{\rm c2}$ as shown in Fig.~\ref{hc2abc}(c).
Namely $H^c_{\rm c2}$ is enhanced there.

The enhancement is estimated as in the same manner as the $H_{\rm c2}$ suppression case:
The $T_{\rm c}$ shift is given by $\Delta T=\kappa M_c(H=12T)=2.7(K/\mu_{\rm B})\times 0.25\mu_{\rm B}=0.65$K.
Substituting $\alpha^c_0=5.7(T/K)$, $\Delta H=\alpha^c_0 \Delta T=3.7$T is obtained,
leading to the enhanced $H^c_{\rm c2}=H^{\rm orb}_{\rm c2}+\Delta H=12T+3.7T=15.7$T,
which is nearly observed value $\sim$15T. Thus the $H_{\rm c}$ enhancement is precisely
consistent with the KS experiments~\cite{ishida3,ishida4}.

\section{Under pressure}

In order to understand the pressure evolution of the multiple phase diagrams in the $H$-$T$ plane,
we apply the above theory of the the $H_{\rm c2}$ suppression and enhancement mechanism,
which turns out to be quite fruitful as seen in the following.
By inspecting the overall evolutions of the multiple phase diagrams in the $H$-$T$ plane~\cite{pressure0,pressure1,pressure2,pressure3,pressure4,pressure}
shown in Fig.~\ref{figa} for $H$$\parallel a$-axis, Fig.~\ref{figb} for $H$$\parallel b$-axis, 
and Fig.~\ref{figc} for $H$$\parallel c$-axis from low to high $P$, we understand that\\
 \noindent
(1) The two phases A$_1$ in high $T$ and A$_2$ in lower $T$ at $H$=0 approaches, coincides, and
interchanges  each other at around $P=0.18$GPa above which the A$_2$ (A$_1$) is the high (low) $T$ phase.\\
 \noindent
(2) In addition to the A$_1$ and A$_2$ phases, the A$_0$ phase corresponding to the $\eta_a$ component
appears in the intermediate pressure region centered at
$P_{\rm TCP}=0.18$GPa, and fades away outside of it.
In particular, since the three phases are almost degenerate at around $P_{\rm TCP}$ 
whose transition temperatures coincide in $H=0$, it is difficult to determine
the precise phase boundaries. The information in hand is not enough to unambiguously 
draw the phase boundary lines there.\\
 \noindent
(3) It is noteworthy  as a whole that with increasing $P$ while in $H$$\parallel$$b$-axis and $c$-axis
the phase diagrams progressively expands toward the $T$-axis and $H$-axis, those for $H$$\parallel$$a$-axis
remains suppressed toward the $H$-axis in spite of $T_{\rm c}$ going up to 3K.

\subsection{Phase diagram evolution for $H$$\parallel$$a$-axis}

\begin{figure}
\begin{center}
\includegraphics[width=14cm]{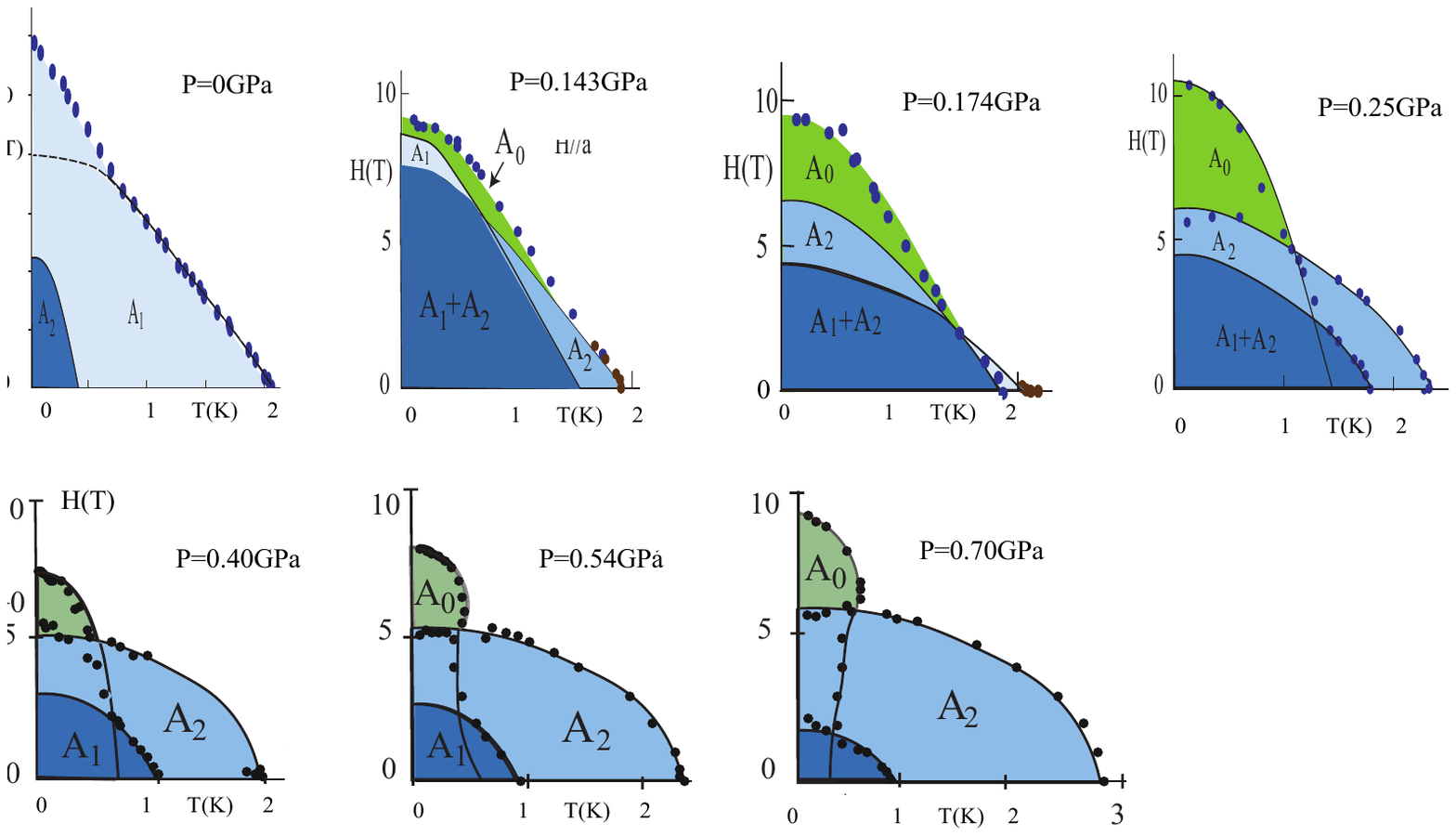}
\end{center}
\caption{\label{figa}
Pressure evolution of the phase diagrams for $H$$\parallel$$a$-axis.
The A$_2$ phase hidden in the low $T$ at the ambient pressure $P$=0 expands toward
the high $T$ and high $H$ directions. Around $P$=0174GPa the transition temperatures
for the two phases coincide at $H=0$-axis, above which the A$_2$ phase becomes the high $T$ phase.
In spite of the growing transition temperature approaching 3K, $H_{\rm c2}$ of the A$_2$ phase
remains largely suppressed around 5T. The extra-high $H$ phase in $P$=0.25GPa, 0.40GPa,
0.54GPa and 0.70GPa is particularly evident and identified as the A$_0$ phase. The data come from Refs.~[\onlinecite{pressure1}], 
 [\onlinecite{pressure2}], and ~[\onlinecite{pressure3}].
}
\end{figure}

We compile all the available data~\cite{pressure1,pressure2,pressure4} on the $T$-$H$ phase diagrams for $H$$\parallel a$-axis in  Fig.~\ref{figa}.
Starting with the ambient pressure toward higher $P$, it is seen that 
the A$_2$ phase progressively manifests itself and occupies larger regions in phase diagrams.
At $P=0.174$GPa which is, we identify, the nearest to the critical pressure $P_{\rm TCP}$ among these figures
the three phases have almost the same transition temperatures at $H$=0.
Away from it in $P=0.25$GPa it becomes clear to see the two transition temperatures $T_{\rm c1}$ 
and $T_{\rm c2}$ separately at $H$=0.
Judging from the extrapolation from the high field data, we can anticipate the lower third transition $T_{\rm c3}$
for the A$_0$ phase, which is not detected experimentally so far. 
Here the highest temperature phase corresponds to the  A$_2$ phase, meaning that this pressure 
is above $P_{\rm TCP}$.
Going further to higher $P=0.40$GPa, $P=0.54$GPa, and $P=0.70$GPa, 
this multiple phase diagram remains essentially the same as seen in Fig.~\ref{figa}.
It is rather remarkable to see that even the transition temperature increases monotonically toward $P=0.70$GPa,
the A$_2$ phase cannot expand to higher field, namely $H_{\rm c2}$ remains strongly suppressed.

The above implies the following: The spin polarization $\bf S$ directed antiparallel to
the $a$-axis, which is the magnetic easy axis, never flips its direction under the 
external field along the $a$-axis. This is physically reasonable that this spin orientation
is a most stable spin-configuration for the system and implemented from the outset.
This is quite different from the other directions $b$ and $c$, whose magnetic energy is
 gained by rotating the spin polarization, or the d-vector rotation.

This implies that the spin polarization $\bf S$ for the A$_2$ phase is antiparallel to the $a$-direction.
This is the same direction as the A$_1$ phase for $P<P_{\rm TCP}$.
That is, the high temperature phase has always the spin polarization $\bf S$ antiparallel to the field 
direction $H$$\parallel$$a$-axis throughout the whole $P$ region. Crossing $P_{\rm TCP}$ does not alter the spin polarization.
This is a bit surprising because the two transition temperatures crosses at  $P_{\rm TCP}$
by keeping the same spin-polarization. We note that the jumps of the specific heat at the
transition temperatures at higher $T$ in $P<P_{\rm TCP}$ are larger than those in lower $T$ while
these are reversed in $P>P_{\rm TCP}$. The A$_1$ phase and the A$_2$ phase are distinctive entities
characterized by having such as different density of states, etc
as a superfluid condensate, yet they have the same spin polarization.
We will investigate its origin later.

We point out that the existence of the A$_0$ phase is evident in this $H$$\parallel$$a$-axis case
because the A$_0$ phase stands up as an extra-high field above the others.
This is compared with the other directions $b$-axis (Fig.~\ref{figb}) and $c$-axis (Fig.~\ref{figc}) cases where there is no or little trace for it in the 
phase diagrams.

\subsection{Phase diagram evolution for $H$$\parallel$$b$-axis}

\begin{figure}
\begin{center}
\includegraphics[width=14cm]{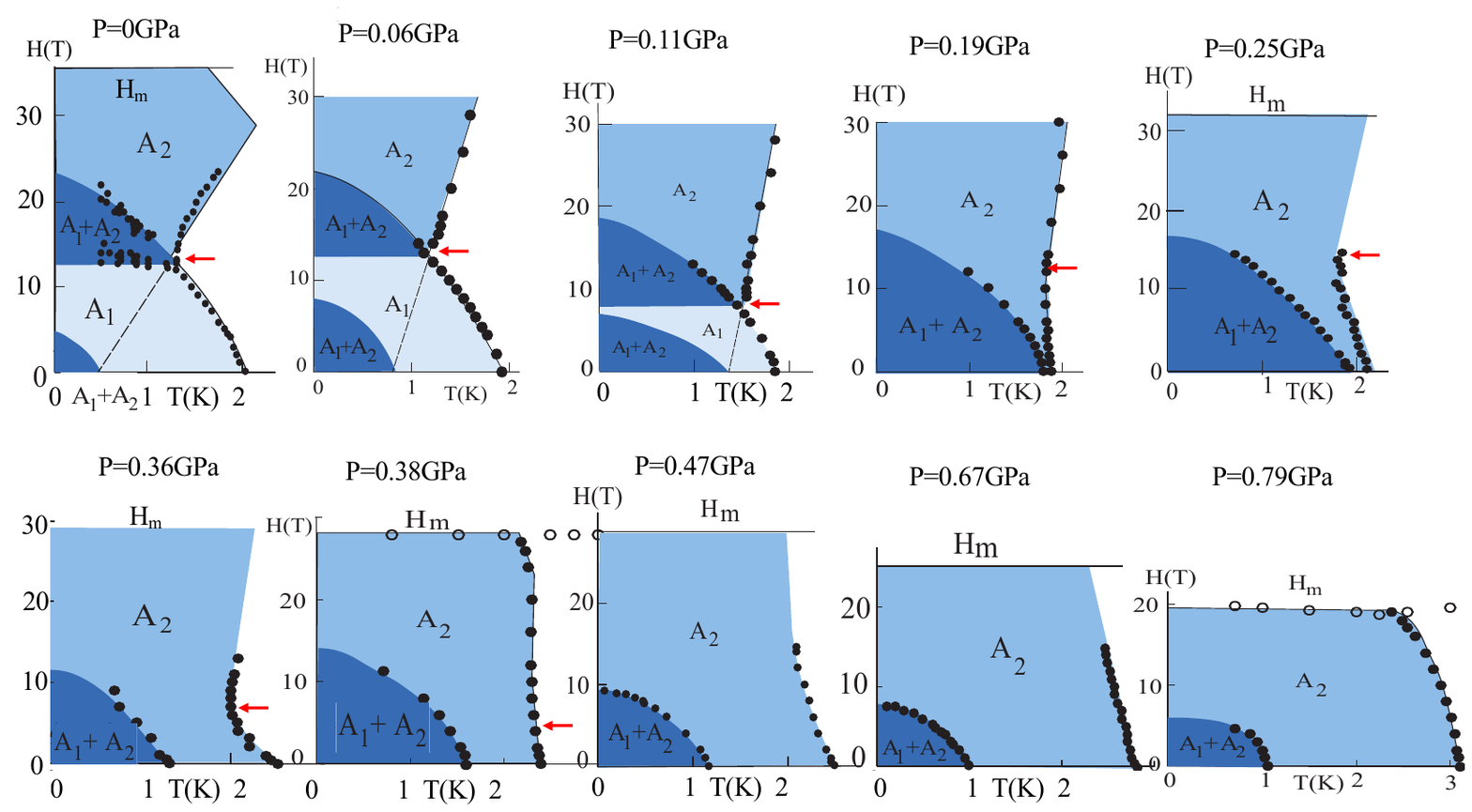}
\end{center}
\caption{\label{figb}
Pressure evolution of the phase diagrams for $H$$\parallel$$b$-axis.
Starting with the ambient pressure $P=0$ phase diagram with the A$_1$ and A$_2$ phases,
the multiple phases are evolving toward the high $P$ up to $P=0.79$GPa.
It is seen that the A$_1$ phase at $P=0$, containing the hidden A$_2$ phase
in the low $T$ and higher $H$ as the intermediate phase is shrinking progressively.
Around $P$=0.19GPa the transition temperatures of the two phases coincide
above which the A$_1$ phase is embedded in the A$_2$ phase.
The positive sloped $H_{\rm c2}$ at $P=0$ associated with the kink structure indicated by the red arrow
becomes weaken in $P=0.06$, and 0.11GPa systematically because of increasing $T_{\rm c2}$.
The fields of the kink position denoted by the red arrows lower.
At $P$=0.19GPa the kink structure reappears around higher field $H$=10T and progressively becomes lower
and disappear. The data points come from Refs.~[\onlinecite{pressure2}], ~[\onlinecite{pressure3}], and ~[\onlinecite{pressure}].
}
\end{figure}

The evolution of the phase diagrams under $P$ in the $b$-axis~\cite{pressure,pressure2,pressure3} is displayed
in Fig.~\ref{figb}. At $P$=0 the A$_1$ phase in low $H$ and the A$_2$ phase in high $H$ with
$T_{\rm c1}$$>$$T_{\rm c2}$
are sandwiched by the intermediate phase, a mixture of A$_1$ and A$_2$ phases denoted as 
A$_1$+A$_2$ in this figure.
By increasing $P$ the A$_2$ phase expands to higher $T$ region and eventually the two
transitions $T_{\rm c1}$ and $T_{\rm c2}$ coincide at $P_{\rm TCP}$ seen at $P=0.19$GPa in Fig.~\ref{figb},
above which $T_{\rm c1}$$<$$T_{\rm c2}$.
Judging from this $P$ evolution, it is natural to postulate that even at lower $P$,
including the ambient pressure in particular, the A$_2$ phase exists at lower $T_{\rm c2}$ at $H=0$.
Then the $P$ evolution is easily understood as the A$_2$ phase in low $T$ and low $H$
evolves simply toward higher $T$ regions. This picture is explained in detail 
in the previous publications~\cite{machida4,machida5}, including the appearance of the
intermediate region A$_1$+A$_2$ and the tetra-critical point indicated by the red arrow in $P=0$.

According to the present scenario, the two transitions
are described by $T_{\rm c1}=T_{\rm c0}+\kappa M^{(0)}_a$
and $T_{\rm c2}=T_{\rm c0}-\kappa M^{(0)}_a$ at $H=0$ where the hypothetical spontaneous moment
$M^{(0)}_a$ is the root mean square average. We attribute the
$P$ evolution to varying the magnitude of $\kappa(P)$, keeping its sign non-positive.
Although it might be possible to their changes due to 
$M^{(0)}_a(P)$ as an alternative, we turn down its possibility 
because it is hard to believe that in the narrow $P$ region around
$P_{\rm TCP}$ the easy axis magnetization $M^{(0)}_a$ drastically varies
from a positive to a negative value through $M^{(0)}_a=0$ at $P_{\rm TCP}$.
In fact the susceptibility measurement~\cite{li} under $P$ 
shows a smooth and little change for all directions: $\chi_a$, $\chi_b$,
and $\chi_c$. A notable change is that the so-called $\chi_b$ maximum
temperature is lower as $P$ increases as evidenced by lowering the metamagnetic 
transition field $H_{\rm m}$.

It shoud be noticed from Fig.~\ref{figb}:\\
\noindent
(1) The positive slopes of $H_{\rm c2}$ seen in $P$=0, 0.05, and 0.11GPa become weak.\\
\noindent
(2) The associated kink positions indicated by the red arrows lower in $H$.\\
\noindent
(3) However, it increases suddenly at $P$=0.19GPa and then lowers again toward high $P$.\\
\noindent
(4) Thus in the higher $P$=0.47, 0.67, and 0.79GPa, $H_{\rm c2}$ strongly increases from $H$=0
with a large slope.\\
\noindent
(5) At the metamagnetic transition $H=H_{\rm m}$, $H_{\rm c2}$ always terminate suddenly.

These items are further investigated later and reveal the physical reasons why it is so.

\subsection{Phase diagram evolution for $H$$\parallel$$c$-axis}

\begin{figure}
\begin{center}
\includegraphics[width=14cm]{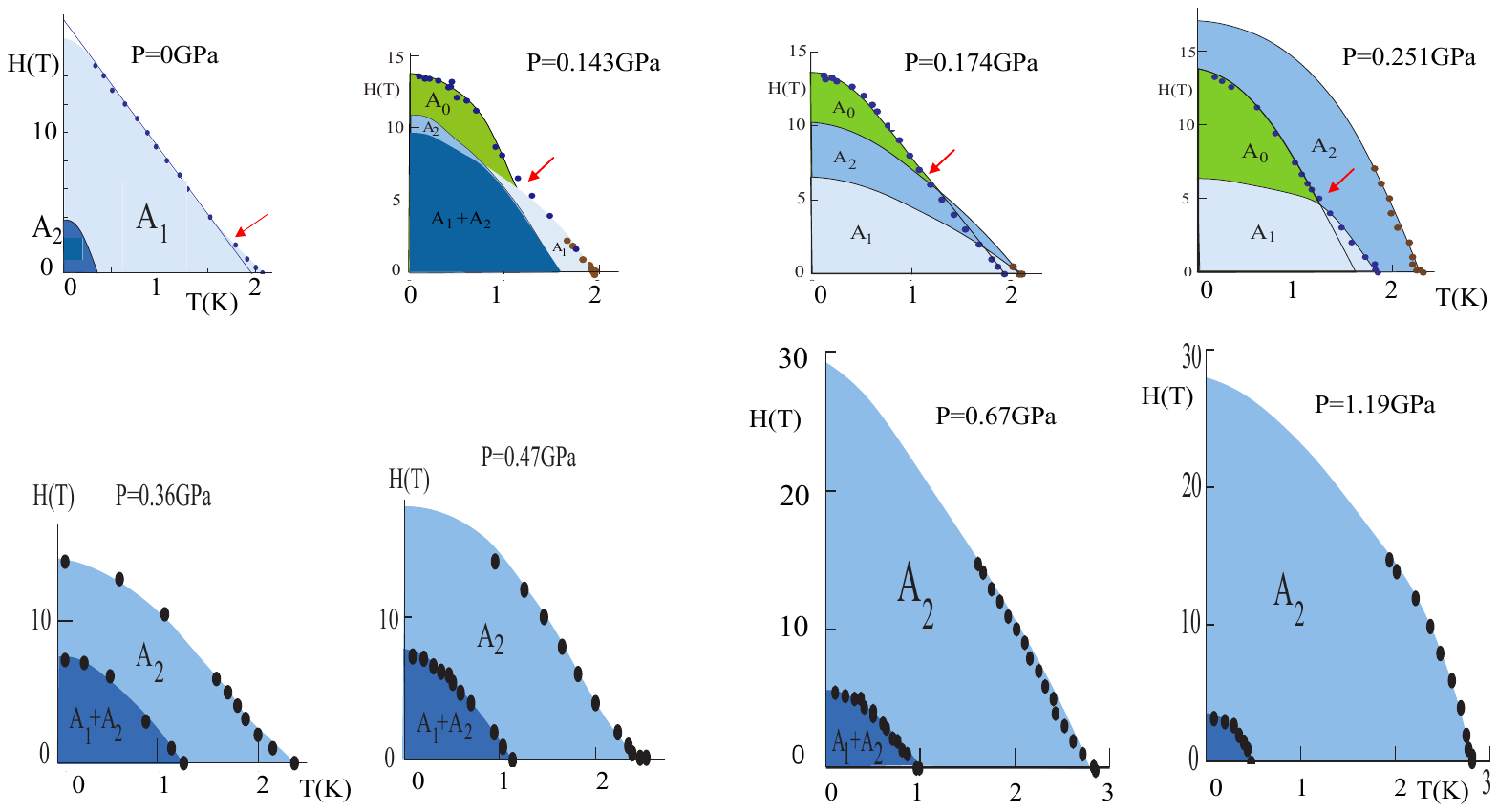}
\end{center}
\caption{\label{figc}
Pressure evolution of the phase diagrams for $H$$\parallel$$c$-axis.
Starting with the phase diagram with the A$_1$ and A$_2$ phases at the ambient pressure $P=0$,
the multiple phases are evolving toward the high $P$ up to $P=1.19$GPa.
The overall change of the two  A$_1$ and A$_2$ phases is to exchange its places in
the $H$-$T$ plane. The two ends at $P=0$ and $P=1.19$GPa are similar.
As $P$ increases, $H^c_{\rm c2}$ expands both toward $H$-direction and $T$-direction.
In $P=0.143$GPa, 0.174GPa, and 0.251GPa the red arrows denote the anomalies,
indicating the existence of the additional third phase A$_0$.
The data come from Refs.~[\onlinecite{pressure2}], ~[\onlinecite{pressure3}], and~[\onlinecite{pressure4}].
}
\end{figure}

Finally, we examine the multiple phase diagrams for $H$$\parallel$$c$-axis~\cite{pressure,pressure4,pressure2}.
Under $P=0$ the kink structure of $H^c_{\rm c2}$ is understood as corresponding
to the d-vector rotation field. Since in the zero field the spin polarization $\bf S$
points antiparallel to the  $a$ direction, this low field rotation continues to be true
throughout all $P$ cases shown in Fig.~\ref{figc}.
It is seen from Fig.~\ref{figc}:\\
\noindent
(1) The high $T$ phase A$_1$ at $P=0$ is simply shrinking their areas with $P$.\\
\noindent
(2) The low $T$ phase A$_2$ at $P=0$ is simply expanding their areas with $P$.\\
\noindent
(3) Thus, $P=1.19$GPa phase diagram looks similar to that in $P=0$ except that
the two phases A$_1$ and A$_2$ exchange its position in $H$-$T$ phase diagrams.\\
\noindent
(4) Toward higher $P$, $H^c_{\rm c2}$ for the A$_2$ phase continues to be larger.
Namely, there is no trace for the $H_{\rm c2}$ suppression, rather we see
the $H^c_{\rm c2}$ enhancement. This is reasonable because the d-vector rotation field
situates at lower $H$ in this axis $c$.\\
\noindent
(5) Although it is subtle to see the A$_0$ phase in $P=0.143$GPa,  and 0. 174GPa
where we see small enhancements of $H^c_{\rm c2}$ denoted by the red arrows,
it is rather clear to see an anomaly in the phase boundary between the  A$_0$ phase
and the A$_1$ phase indicated by the red arrow in $P=0.251$GPa. These anomalies correspond to the 
A$_0$ phase.

\section{Origin of the pressure evolution of the multiple phase diagrams}

We are now in position to investigate the origin why the multiple phases evolve under $P$.
As we point out above that the underlying magnetic system hardly
changes throughout the pressure region of interest~\cite{li}. The governing factor to yield the
pressure evolution of the multiple phase diagrams is something other than that, which we investigate now.

\subsection{Pressure dependence of $H_{\rm rot}$}

\begin{figure}
\begin{center}
\includegraphics[width=10cm]{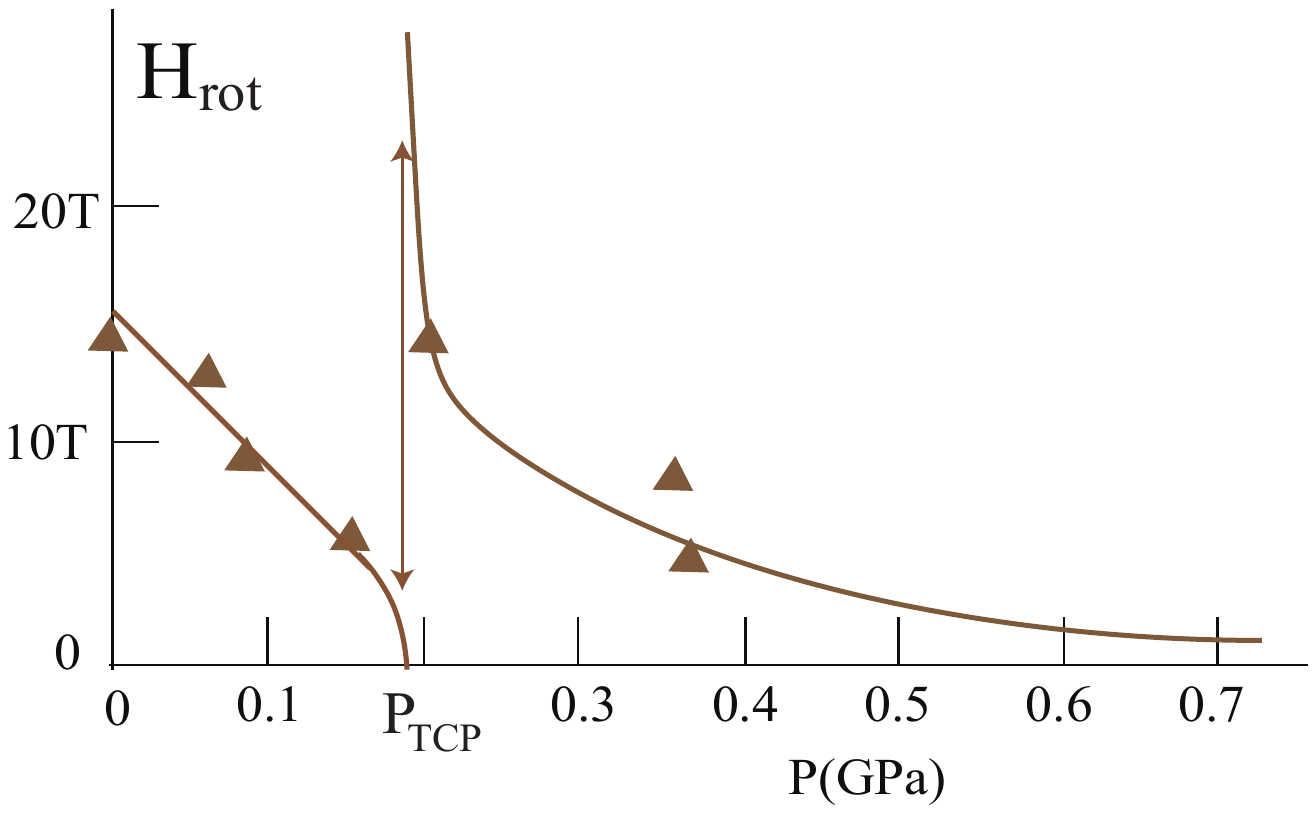 }
\end{center}
\caption{\label{rot}
$P$ dependence of the d-vector rotation field $H_{\rm rot}(P)$
for $H$$\parallel b$-axis, extracted from Fig.~\ref{figb}.
$H_{\rm rot}(P)\rightarrow\pm\infty$ toward $P_{\rm TCP}$ from the both sides.
}
\end{figure}

Let us examine  the $P$ dependence of $H_{\rm rot}$ for $H$$\parallel$$b$-axis plotted in Fig.~\ref{rot}, which 
is extracted from Fig.~\ref{figb}. At $P_{\rm TCP}$, the d-vector rotation field $H_{\rm rot}(P)$ exhibits a jump: From the lower $P$ side
$H_{\rm rot}(P)$ becomes quickly to lower fields while from the higher $P$ side toward $P_{\rm TCP}$
it becomes larger.
The former $P$ dependence is attributed to the fact that $T_{\rm c2}(P)$ for the hidden A$_2$ phase 
situated with the lower $T$ region
increases quickly toward $T_{\rm c1}(P)$, which is relatively unchanged in this $P$ region.
Therefore, $H_{\rm rot}(P)$ which corresponds to the tetra-critical point in the $H$-$T$ plane moves down to lower fields.
On the other hand, the latter behavior for $P>P_{\rm TCP}$ can be understood in terms of the competition between the 
spin-orbit coupling energy $E_{\rm SOC}$,
which acts as the locking $\bf S$ to the crystalline lattices, and the magnetic energy coming from the $\kappa$-term
in the GL functional, or $\kappa M(H)=\kappa\chi H$. By equalizing the both terms: 
$\kappa\chi H_{\rm rot}$=$E_{\rm SOC}$,
we find $H_{\rm rot}\propto 1/\kappa$ under the assumption that $E_{\rm SOC}$ is insensitive of $P$
in this narrow pressure region around $P_{\rm TCP}$. 
This means that when approaching from the high (low) $P$ side to $P_{\rm TCP}$,
$H_{\rm rot}\rightarrow \pm\infty$ as shown in Fig.~\ref{rot}.

\subsection{Pressure dependence of $\kappa(P)$}

It is obvious to see that $\kappa(P)$  linearly changes in $P$  away from $P_{\rm TCP}$, 
namely $\kappa(P)\propto |P_{\rm TCP}-P|$
because $|T_{\rm c1}-T_{\rm c2}|=\kappa M^{(0)}_a$ at $H$=0 where $|T_{\rm c1}-T_{\rm c2}|$ is linear in $P$ near $P_{\rm TCP}$.
Here we assume that $M^{(0)}_a$ is independent of $P$ around $P_{\rm TCP}$.

\begin{figure}
\begin{center}
\includegraphics[width=10cm]{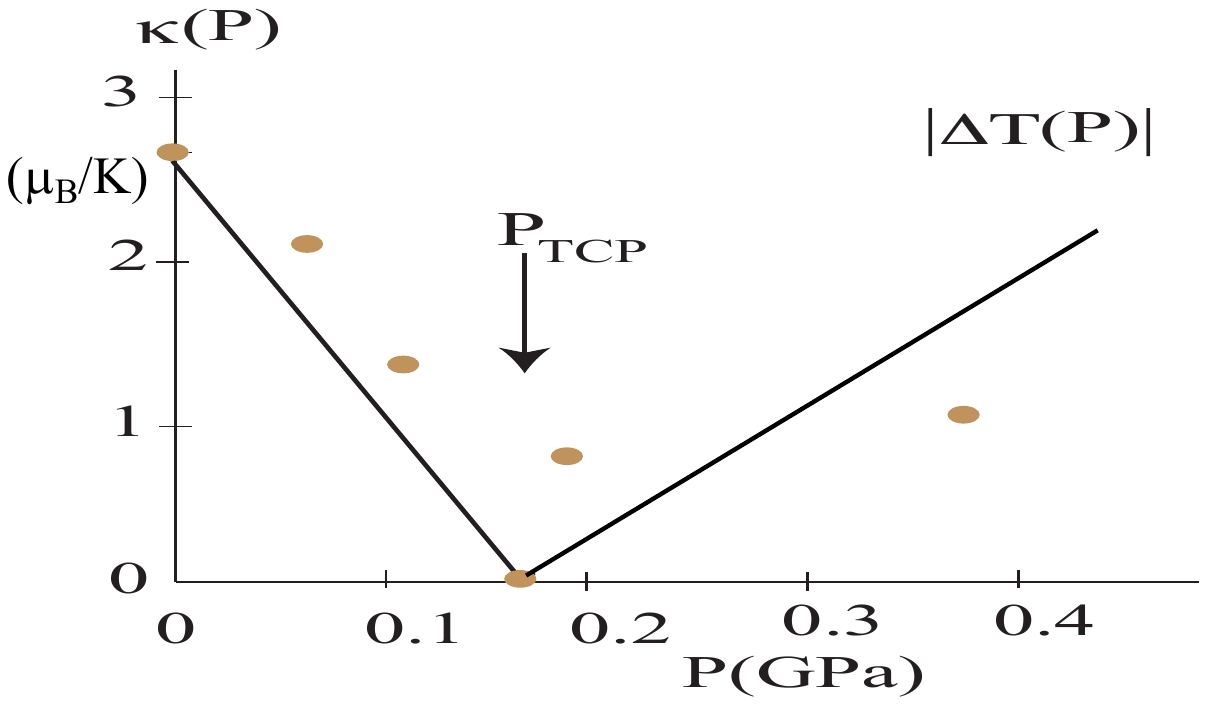}
\end{center}
\caption{\label{kappa}
$P$ variation of $\kappa(P)$ (the dots) estimated by the slopes in Fig.~\ref{hc2abc}(b)
where $\kappa(P)\propto \Delta M_b(H=30T)$. The straight lines indicate $\kappa(P)$ estimated from
$|\Delta T|\equiv|T_{\rm c1}-T_{\rm c2}|=\kappa M^{(0)}_a$. Note that $\kappa(P_{\rm TCP})$=0.
}
\end{figure}

In order to check the pressure dependence of $\kappa(P)$, we examine the positive slopes of  $H_{\rm c2}$ shown in 
Fig.~\ref{figb} because the slope is determined by $\kappa M_b(H)$ as discussed in Fig.~\ref{hc2abc}(b). 
We can extract the relative $\kappa$ values for $P$=0.06, 0.11, and 0.19GPa to $\kappa$=2.7(K$/\mu_{\rm B})$ 
at $P=0$ from Fig.~\ref{figb} by measuring $\Delta T=\kappa M_b$ at $H$=30T where $M_b$ is assumed to be unchanged.
 The results in Fig.~\ref{kappa} show that the $\kappa(P)$ values systematically decrease with $P$ from $P=0$ toward $P_{\rm TCP}$.
Then, after passing $P_{\rm TCP}$ where $\kappa(P_{\rm TCP})$=0, it increases again to larger values.
This tendency qualitatively matches with the variation $\kappa(P)$ extracted $|T_{\rm c1}-T_{\rm c2}|=\kappa M^{(0)}_a$ denoted by the 
straight lines as  $|\Delta T(P)|$.

\subsection{$P$ phase diagram and possible origin of $\kappa(P)$}

We first recall the expression~\cite{mermin} for 

$$\kappa=T_{\rm c}{N'(0)\over N(0)}ln(1.14\Omega_{\rm c}/T_{\rm c}).$$

\begin{figure}
\begin{center}
\includegraphics[width=12cm]{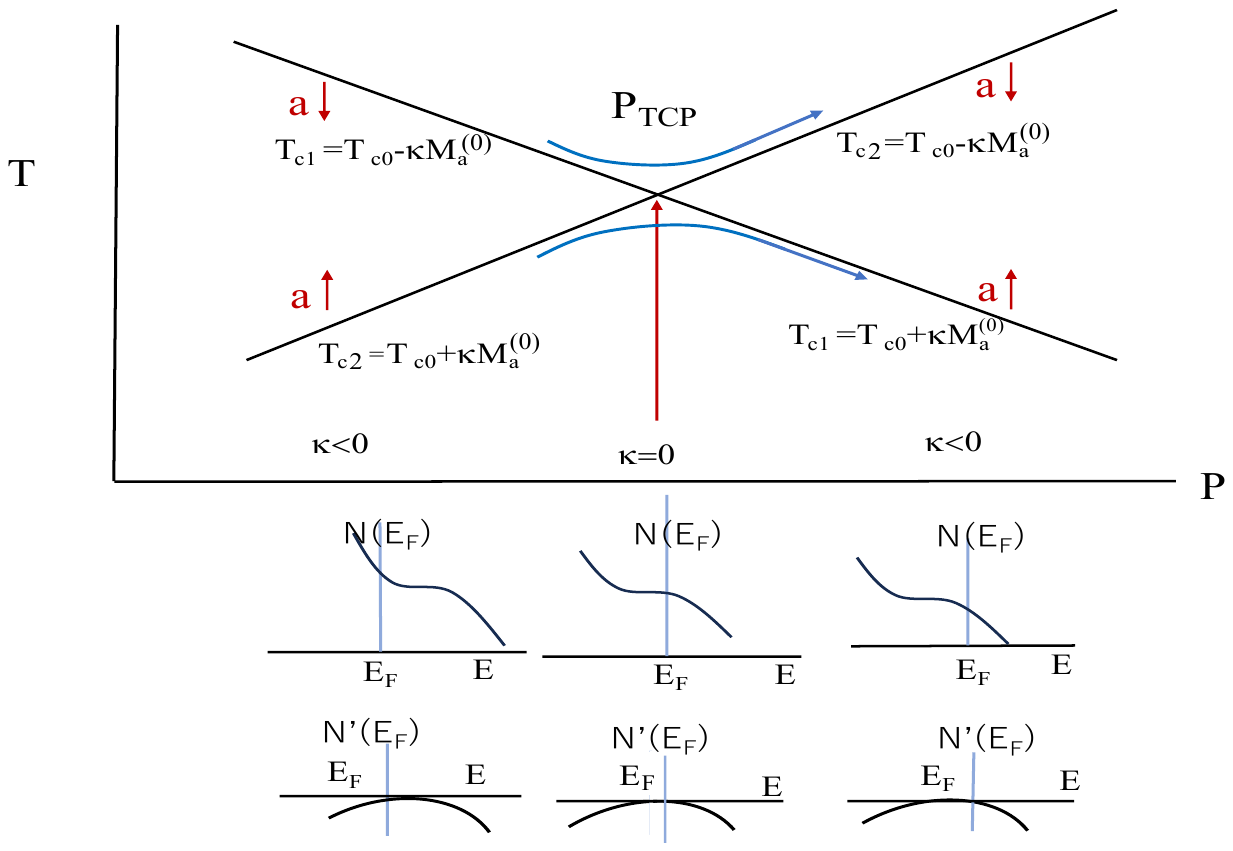}
\end{center}
\caption{\label{Pphase}
Schematic phase diagram in $T$ and $P$ plane.
In the low $P$ side the A$_1$ (A$_2$) phase at high (low) $T$
has the transition temperature $T_{\rm c1}=T_{\rm c0}-\kappa M^{(0)}_a$ ($T_{\rm c2}=T_{\rm c0}+\kappa M^{(0)}_a$)
is characterized by the spin polarization $\bf S$ pointing antiparallel (parallel) to the $a$-direction.
In the high $P$ side the A$_2$ (A$_1$) phase at high (low) $T$
has the transition temperature $T_{\rm c1}=T_{\rm c0}-\kappa M^{(0)}_a$ ($T_{\rm c2}=T_{\rm c0}+\kappa M^{(0)}_a$)
is characterized by the spin polarization $\bf S$ pointing antiparallel (parallel) to the $a$-direction.
The two transition temperatures meet at $P_{\rm TCP}$.
Throughout $P$ region $\kappa\le0$.
The two lows in the bottom show the Fermi level $E_F$ shifts in the DOS  $N(E_F)$ and its derivative $N'(E_F)$ under $P$.
}
\end{figure}

\noindent
The energy derivative $N'(E_F)$ of DOS $N(E_F)$ at the Fermi level $E_F$(=0) can be zero
when $N(E_F)$ becomes either extreme, such as a maximum and  minimum or an inflection point.
In the former case $\kappa(P)$ changes its sign around the extreme while
in the latter case $\kappa(P)$ keeps the same sign around the inflection point.
Therefore as a possibility if  $N(E_F)$ is a decreasing function of $E_F$ with an
inflection point as shown in the bottom low of Fig.~\ref{Pphase}, it may explain the $P$ variation of 
$\kappa(P)$ under the assumption that $E_F(P)$ shifts from the left to right in the energy $E$
axis under $P$ where $P_{\rm TCP}$ corresponds to the inflection point with $\kappa(P_{\rm TCP})=0$. 
Thus $\kappa(P)\le0$ is kept throughout the entire $P$ region, consistent with our picture shown in Fig.~\ref{kappa}
and the discussions in Sec. IV.

As shown in Fig.~\ref{Pphase} as a schematic diagram in the $T$-$P$ plane at zero field,
we can assign the spin polarization $\bf S$ with their direction and up-down orientations
for each phase where we suppress the A$_0$  phase for clarity. Here we restore the notation, $\kappa<0$:\\
\noindent
(1) For $P<P_{\rm TCP}$:
The Cooper pair spin $\bf S$ polarizes along the $a$-axis with $\downarrow\downarrow$ ($\uparrow\uparrow$) pairs in the
high (low) $T$ phase A$_1$ (A$_2$) of $T_{\rm c1}=T_{\rm c0}-\kappa M^{(0)}_a$ ($T_{\rm c2}=T_{\rm c0}+\kappa M^{(0)}_a$).\\
\noindent
(2) For $P=P_{\rm TCP}$:  At the tetra-critical point where the four second order phase transition lines meet
and reconnected guided by the arrows there. This critical point is akin to the TCP in $H$$\parallel$$b$-axis in ambient pressure. 
Here $\kappa$=0, corresponding to the inflection point in DOS shown in the lowest lows in Fig.~\ref{Pphase}. We assumed
that the electron density is kept constant by modifying the overall band structure.
\\
\noindent
(3) For $P>P_{\rm TCP}$: In the
high (low) $T$ phase A$_2$ (A$_1$) of $T_{\rm c1}=T_{\rm c0}-\kappa M^{(0)}_a$ ($T_{\rm c2}=T_{\rm c0}+\kappa M^{(0)}_a$).
But the high $T$ phase A$_2$ (A$_1$) is characterized by the Cooper pairs polarization
$\bf S$$\parallel$$a$-axis with $\downarrow\downarrow$ ($\uparrow\uparrow$) pairs.
$\kappa\le0$ is kept always.
Therefore, the KS drops always when entering the SC from the normal state
at $H$=0 or in lower fields. 

Note that according to Kinjo et al~\cite{kinjo} who perform
the KS experiment at $P=1.2$GPa for $H$$\parallel$$b$-axis for $H$=0.8T, 1.0T, and 2.5T.
The results show that at $T_{\rm c2} (>T_{\rm c1})$ the KS remains the normal value and drops
at $T_{\rm c1}$. This can be understood because as mentioned above shown in Fig.~\ref{rot} 
the d-vector rotation field becomes low and their measurements senses the spin polarization flipped
along the $b$-direction to save the magnetic energy.

\begin{figure}
\begin{center}
\includegraphics[width=12cm]{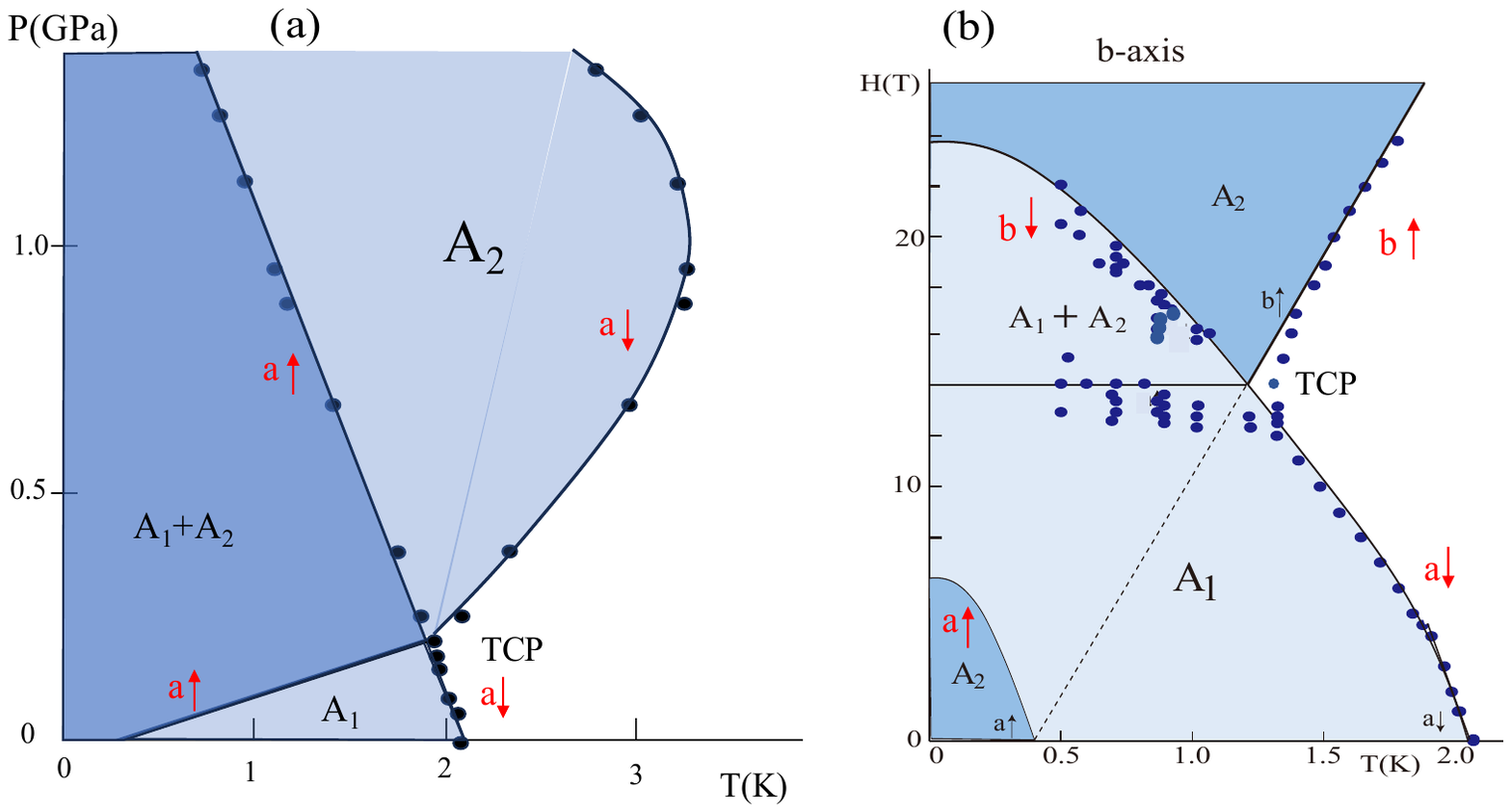}
\end{center}
\caption{\label{comparison}
(a) $P$ vs $T$ phase diagram where the A$_0$ phase is omitted for clarity.
Each phase is characterized by the spin polarization and its direction
where the spin quantization axis is along $a$ for all phases.
(b) $H$ vs $T$ phase diagram for $H$$\parallel$$b$-axis.
Each phase is characterized by the spin polarization and its direction
where the low $H$ the spin quantization axis is along $a$ while it is along $b$
in the higher $H$ above  the tetra-critical point (TCP).
}
\end{figure}

In this respect, it might be useful to compare the phase diagrams of $P$-$T$ plane in Fig.~\ref{comparison}(a) and $T$-$H$ plane
in ambient pressure of Fig.~\ref{comparison}(b) to see the different roles played by $P$ and $H$
although they look similar.
It is seen from Fig.~\ref{comparison}(a), the spin polarization $\bf S$ always points to either antiparallel
or parallel to the $a$-direction because there is no reason energetically to change its direction
under $P$. Only $\kappa(P)$ evolves, keeping its sign to be negative. $P_{\rm TCP}$ signifies
the point at $\kappa(P_{\rm TCP})$=0 where $T_{\rm c1}$=$T_{\rm c2}$ (=$T_{\rm c3}$, not shown).

On the other hand, in the $T$-$H$ plane (see Fig.~\ref{comparison}(b)),
starting with $\bf S$ antiparallel or parallel to the $a$-axis at the low $H$,
TCP signifies the d-vector field at $H=14$T, above which $\bf S$ turns to the
$b$-direction to save the magnetic energy associated with the $\kappa$-term in GL
functional.
This is fully reasonable because the magnetization $M_b(H)$ becomes larger with $H$
and the Cooper pairs take advantage of its condensation energy by flipping the
spin polarization direction when $\kappa M(H)>E_{\rm SOC}$.

\begin{figure}
\begin{center}
\includegraphics[width=12cm]{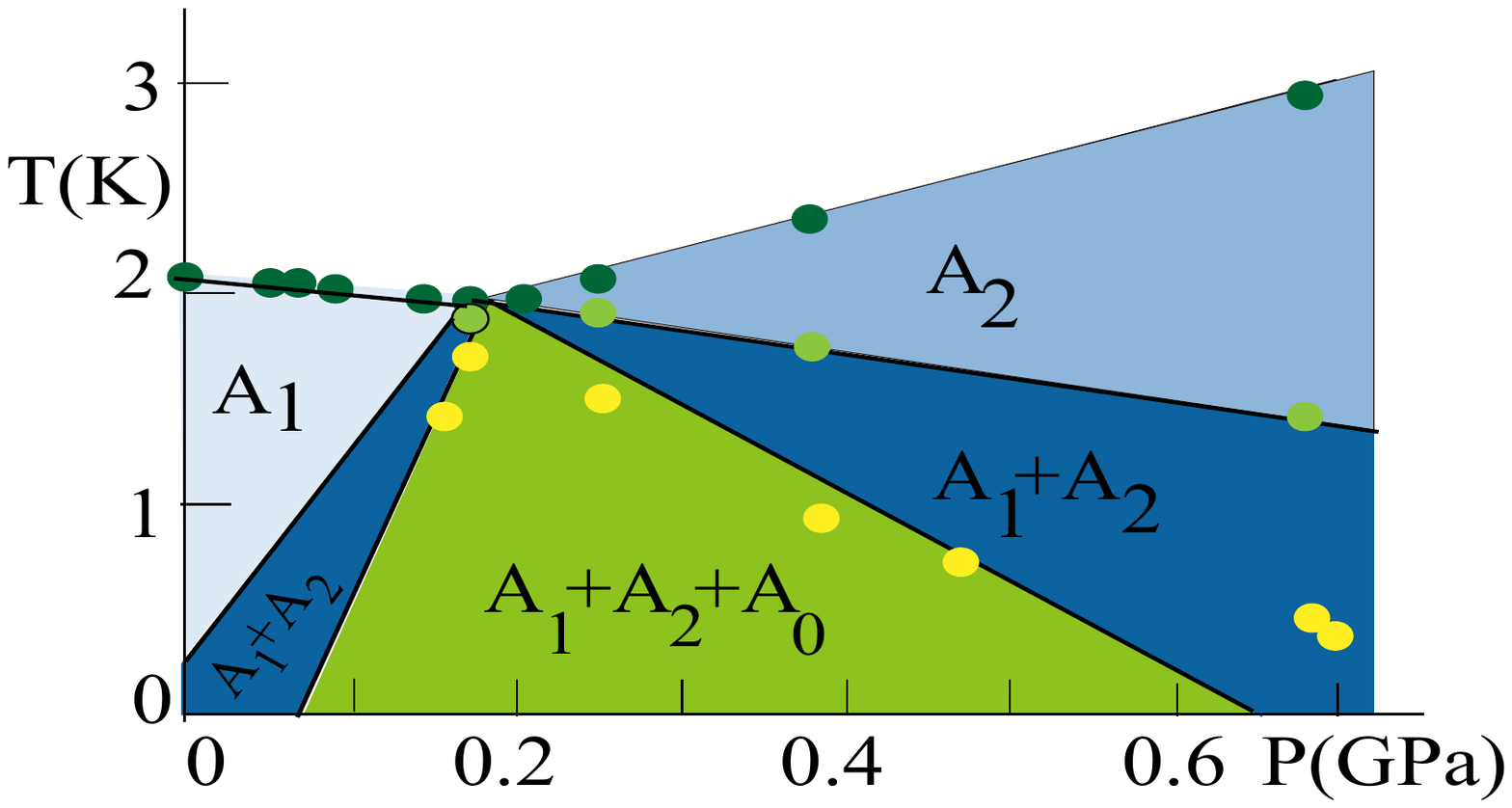}
\end{center}
\caption{\label{A1A2A0}
The detailed phase diagram in $T$-$P$ plane with the
A$_1$, A$_2$, and A$_0$ phases. The dots are experimental points
coming from Aoki et al~\cite{pressure0,pressure1,pressure2,pressure3,pressure4}. The yellow dots are estimated from 
data shown in Figs.~\ref{figa},~\ref{figb}, and~\ref{figc}.
}
\end{figure}

The more accurate $P$-$T$ phase diagram is displayed in Fig.~\ref{A1A2A0}
where the A$_0$ phase is estimated from Figs.~\ref{figa},~\ref{figb}, and~\ref{figc}.
It is seen from Fig.~\ref{A1A2A0} that\\
\noindent
(1) In the lower $P$ side, the A$_1$ phase is the high $T$ phase. At lower $T$
the A$_2$ phase appears via a second order phase transition.\\
\noindent
(2) In the high $P$ side, the A$_2$ is the high $T$ phase. At lower $T$
the A$_1$ phase appears via a second order phase transition.\\
\noindent
(3) They meet at the tetra-critical point $P_{\rm TCP}$$\sim$0.18GPa.\\
\noindent
(4) At further low $T$, the A$_0$ phase as the coexistence state with A$_1$ and A$_2$
appears centered around $P_{\rm TCP}$, whose complicated
phase boundary structure is not known theoretically and experimentally.\\
\noindent
(5) Generally the lower $T$ phases are a mixture of their phases.
However, the phase with A$_1$+A$_2$, is not identical to the so-called A phase in the
superfluid $^3$He and also the phase with A$_1$+A$_2$+A$_0$, is not identical to the so-called B phase in the
superfluid $^3$He because their transition temperatures are different and they are the distorted A and B phases
in the superfluid terminology~\cite{3he}.

\section{Discussion}

\subsection{Origin of the double transition; spin vs orbital degeneracy scenarios}

Based on the successful analyses in this paper, we conclude that 
the pairing symmetry realized in UTe$_2$ should be a spin triplet state
whose spin part belongs to the equal spin states consisting of the $\uparrow\uparrow$ 
and $\downarrow\downarrow$ pairs as the A$_1$ and A$_2$ phases.
These A-like states are quite versatile in explaining and understanding a variety of 
experimental facts compiled so far~\cite{review}, including the $H_{\rm c2}$ suppression and enhancement
and multiple phases observed in this material.

A possible alternative scenario within the spin triplet pairings may be
that the degeneracy comes from the orbital part of the pairing function,
which explains the multiple phases due to accidental degeneracy of two irreducible representations~\cite{ishizuka1,ishizuka2}
because in the present orthorhombic symmetry only the one dimensional representations are
present with different transition temperatures in the infinitely strong SOC classification. This seems an unsatisfactory scenario
from various unlikely aspects, in particular, the gradual d-vector rotation phenomena observed in $H$$\parallel$$b$-axis
and the observed tetra-critical point with the two second order phase transition lines without ``level repulsion''.

\subsection{SOC and classification scheme}

Our scenario is based on the group theoretical classification scheme with finite spin orbit coupling (SOC)~\cite{machida0,ozaki1,ozaki2,annett}.
Proposed theories~\cite{review}, including accidental degeneracy scenario~\cite{ishizuka1,ishizuka2} usually assumes that the SOC is infinitely strong~\cite{review}, thus in classifying it
the spin and orbital degrees of freedom are tightly coupled and transform together under the group symmetry actions. This infinite strong
SOC scheme is originated long ago~\cite{anderson,blount,gorkov}.
In this infinite SOC the Cooper pair spin is locked to the underlying lattice and never gives rise to the d-vector rotation under an external
field. The controversy over either  finite SOC or infinite SOC starts from the beginning of the discovery of heavy Fermion superconductors, such as 
U$_{1-x}$Th$_x$Be$_{13}$~\cite{ott,shimizu11,shimizu12,shimizu13,Th,Th1} with multiple phases. 
It is acute particularly in UPt$_3$ concerning the existence of the tetra-critical point of
 the multiple phase diagram in the $T$-$H$ plane because according to the scenario on infinite SOC there is no true TCP in general
 because the so-called gradient coupling washes out TCP by the level repulsion term in the GL~\cite{sauls,upt3,ohmi}.
 Namely, the two intersecting second order transition lines are avoided.
 Since the d-vector rotation is observed in UPt$_3$~\cite{tou1,tou2}, the finite SOC scenario is more favorable and infinite SOC is not appropriate.
 According to our finite SOC theory, the gradual d-vector rotation 
 quite possible because the d-vector rotation is controlled by the competition
 between the SOC which locks the d-vector to crystal lattices and the magnetic energy.
 Thus depending on the strengths of the two factors, the rotation occurs gradually 
 at finite fields. It is desired to calculate the strength of the SOC in UTe$_2$ by a microscopic theory~\cite{yanase}
 in light of the estimated SOC coupling constants: $\sim$1T for the $c$-axis and 5T$\sim$14T for the $b$-axis. 
 
\subsection{Pairing symmetry and nodal structure}

According to the finite SOC scheme, the classified pairing functions are all characterized by a line node~\cite{annett}.
This is in stark contrast with these in the infinite SOC scheme where all basis functions classified in D$_{\rm 2h}$ are characterized
by a point node~\cite{review}, since the Blount theorem forbids a line node in this scheme~\cite{blount}
except for known cases~\cite{norman,sato}.
As for the nodal structure in UTe$_2$ it still remains unsettled, ranging from a point node~\cite{metz,kittaka,shibauchi,hayes,roman} to a full gap~\cite{matsuda}. 
Here the nodal structure with a line node is a generic feature in the present scenario.
According to the recent angle-resolved specific heat experiment and theoretical analysis
supports this nodal structure~\cite{kittaka2}.
Therefore, the pairing function $(\hat{b}+i\hat{c})$$k_a$, which is known as the so-called $\beta$ phase in the superfluid $^3$He~\cite{3he}, 
is the most possible symmetry realized in UTe$_2$ at present.
This form is consistent with Theuss et al~\cite{ramshaw} who conclude single component pairing function
in the orbital space.

\subsection{Predictions and possible future experiments}

Here we propose several experiments to check our scenario:\\
\noindent
(1) The Knight shift experiments~\cite{ishida1,ishida2,ishida3,ishida4,ishida5,kinjo,matsumura,kitagawa} are one of the most important  and indispensable methods to
know the structure of the d-vector. We predict that the d-vector does not rotate for the field directions
exactly parallel or antiparallel to the magnetic easy axis $a$.
This is because this particular  d-vector configurations are most stable,
thus to change these stable structures, the magnetic field is needed to be comparable to the
superconducting condensation energy, namely comparable to $H^a_{\rm c2}$.\\
\noindent
(2) Since in $H=0$ and lower $H$ at $P>P_{\rm TCP}$ the spin polarization $\bf S$ points antiparallel to
the $a$-axis for the high $T$ phase, KS should drop below $T_{\rm c2}$($>T_{\rm c1}$).
The existing experiment by Kinjo et al~\cite{kinjo} at $H\ge0.8$T ($H$$\parallel$$b$-axis) under $P$=1.3GPa exhibits to remain unchanged
below $T_{\rm c2}$=3K and drops further lower $T$ at $T_{\rm c1}$=0.5K. 
This is understood that $H=0.8$T$>H_{\rm rot}$ as shown in Fig.~\ref{rot}. It is desired to perform the KS experiments
in lower $H$.\\  
\noindent
(3) The A$_2$ phase at the ambient pressure without a field below $T\sim0.3$K is postulated  in the paper.
This low $T$ phase is similar in their physical properties to the intermediate phase A$_1+$A$_2$ above $H(\parallel b)$=14T.
Thus the ac susceptibility $\chi_{\rm ac}$ or flux flow experiments may detect it as done for the intermediate phase~\cite{sakai}.
We point out that the recent $T_1$ measurement by NMR~\cite{matsumura2} indicates an anomaly at lower $T$,
suggesting unknown phenomenon, possibly the A$_2$ phase.\\
\noindent
(4) The A$_1$ and A$_2$ phases breaks time reversal symmetry, which should be detected by appropriate experimental
methods. The $\mu$SR measurement may not be sufficient because the results are conflicting~\cite{kaptulnik,sonier1,ajeesh}.\\
\noindent
(5) We need more detailed experiments under pressure near $P=P_{\rm TCP}$ to establish the
phase boundaries for the A$_1$, A$_2$, and  A$_0$, in particular, for $H$$\parallel$$a$-axis and $c$-axis.
At $P=P_{\rm TCP}$ the most symmetric state with $T_{\rm c1}$=$T_{\rm c2}$=$T_{\rm c0}$ is realized
described by ${\hat b}k_a$, which is called the polar phase in the superfluid $^3$He~\cite{3he}.

\subsection{Requirements for observing the $H_{\rm c2}$ suppression and enhancement}

The required conditions for this novel mechanism to observe in a spin triplet superconductor with an equal spin pairs are followings:\\
\noindent
(1) The DOS $N(0)$ is particle-hole asymmetric at the Fermi level.\\
\noindent
(2) Its derivative $N'(0)$ with respect to the energy is appreciable.\\
\noindent
(3) The induced moment $M(H)$ by a field should be large.\\
\noindent
These requirements are easily met for heavy Fermion superconductors, such as UTe$_2$
because the quasi-particle DOS for the Kondo systems is a narrow width comparable to the Kondo temperature, thus 
DOS can be asymmetric around $E_F$. The localized 5f electron moments are
large compared with the usual Pauli paramagnetic moment,
the former is an order of 0.1$\mu_{\rm B}$  while the
latter 0.003$\mu_{\rm B}$ for $N(0)$=120mJ/mol K$^2$  at $H$=1T in UTe$_2$.
Moreover, $\kappa\propto N'(0)$ is enhanced by an factor of $E_F/T_{\rm Kondo}$
with $T_{\rm Kondo}$ the Kondo temperature.
Thus the $H_{\rm c2}$ suppression mechanism is generically possible for a spin triplet superconductor,
but it is understood that  the heavy Fermion materials are best suited for its observation.

\subsection{$\gamma(H//a)$}

\begin{figure}
\begin{center}
\includegraphics[width=12cm]{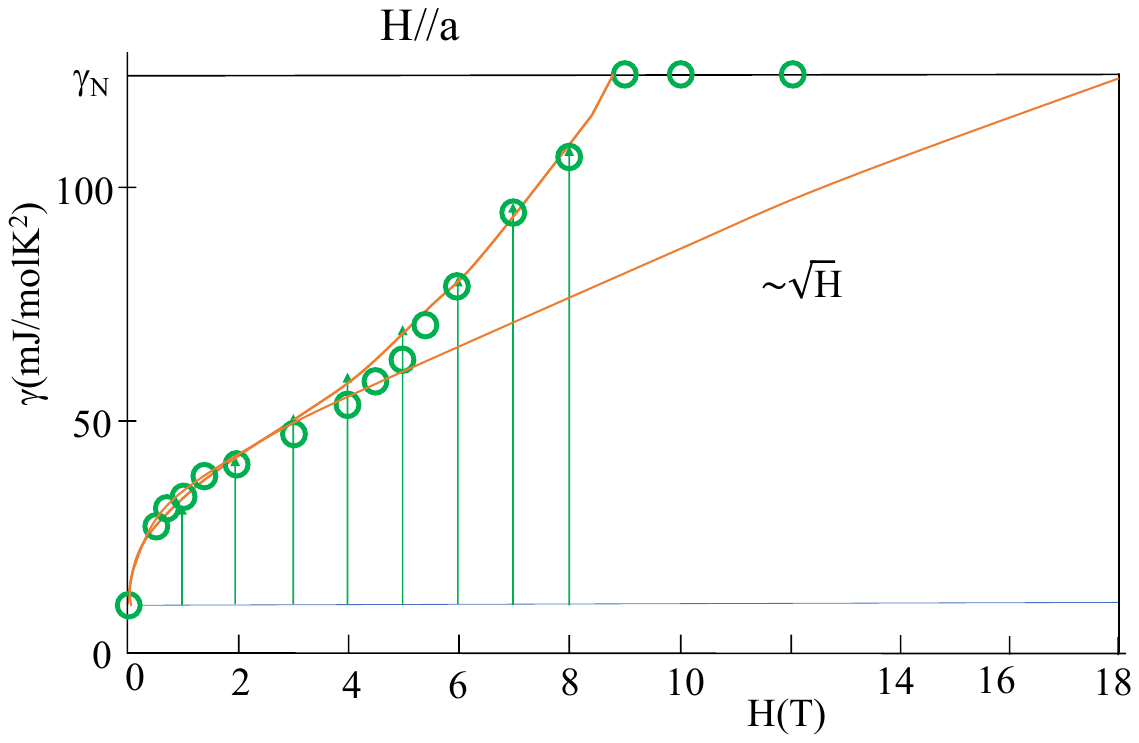}
\end{center}
\caption{\label{gamma(H)}
The comparison with  the theoretical curve and the $\gamma(H)$ data (dots) for the $a$-axis from
 the experiment by Lee, et al~\cite{roman}. The other curve indicates the idealized $\gamma(H)\sim \sqrt H$.
}
\end{figure}

In order to further confirm our assertion on the $H_{\rm c2}$ suppression mechanism realized in UTe$_2$,
we analyze the data of the field evolution of the DOS, namely $\gamma(H)$ for the $a$-axis.
As seen from Fig.~\ref{gamma(H)},  $\gamma(H)$ rises strongly at lower $H$ fitted by $\sqrt H$ like manner
signaling the nodal gap structure.
However, it quickly deviates from $\sqrt H$ behavior and increases further, reaching the its normal value $\gamma_{\rm N}$
at $H^a_{\rm c2}$$\sim$10T, which is far lower than that extrapolated from the initial $\sqrt H$ behavior reached at $\sim$18T.
This coincides with the previous discussion on  the $H_{\rm c2}$ suppression for the $a$-axis.
Thus to understand $\gamma(H)$, we need to take into account of this $H_{\rm c2}$ suppression effect.
In general the nodal gap structure case $\gamma(H)$ is given by the formula
$\gamma(H)/ \gamma_{\rm N}=\sqrt{H/ H_{\rm c2}}$.
Since at $T=0$, $H_{\rm c2}$ is reduced by the magnetization $M(H)$ such that
$H_{\rm c2}=H^{\rm orb}_{\rm c2}-\alpha\kappa M(H)$,
we obtain a formula to evaluate $\gamma(H)$:

\begin{eqnarray}
{\gamma(H)\over \gamma_{\rm N}}=\sqrt{{H\over {H^{\rm orb}_{\rm c2}-\alpha_0\kappa M(H)}}}.
\label{gamma}
\end{eqnarray}

\noindent
After substituting the values known for $H^{\rm orb}_{\rm c2, \parallel a}=30$T,
$\alpha^a_0$=15T, $\kappa=2.7$K/$\mu_{\rm B}$ and $M_a(H)$ shown previously
for the $a$-axis (see Fig.~\ref{deltah}(b)),
we obtain the curve shown in Fig.~\ref{gamma(H)}.

It is seen that as $H$ increases, upon progressively growing $M(H)$, 
$H_{\rm c2}$ is reduced, leading to the rapid growth of $\gamma(H)$.
This curve looks similar to the case~\cite{ichioka1,ichioka2} in the Pauli limited $\gamma(H)$,
which shows a first order transition when the Maki parameter is large.

\subsection{SANS}

In order to see the novel suppression mechanism of the $H_{\rm c2}$ directly,
small angle neutron scattering (SANS) experiments may be a good way to check it.
We start with $H_{\rm c2}(T)=H^{ \rm orb}_{\rm c2}(T)/ {(1+\alpha^a_0\kappa\chi_a})$ 
which is valid for the lower fields with $M_i=\chi_i H$ where $i=a,b$ and $c$.
This means that the vortex unit cell area $S$ compared with $S_0$ in the ordinary superconductors
is reduced by the factor $1+\alpha_0\kappa\chi$, namely

\begin{eqnarray}
{S\over S_0}={1\over {1+\alpha_0\kappa\chi}},
\label{closeforma}
\end{eqnarray}
 
 \noindent
or the unit cell length $L$ of vortex lattices is reduced by

\begin{eqnarray}
{L\over L_0}={1\over \sqrt{1+\alpha_0\kappa\chi}}.
\label{closeforma}
\end{eqnarray}

\noindent
For example, for the $a$-axis 
$H^a_{\rm c2}(T)=H^{ \rm orb}_{\rm c2, {\parallel a}}(T)/(1+\alpha^a_0\kappa\chi_a)$,
where $\alpha^a_0$=15T/K, $\kappa$=2.7K/$\mu_{\rm B}$ and $\chi_a$=0.075$\mu_{\rm B}$/T,
leading to $H^a_{\rm c2}/{H^{ \rm orb}_{\rm c2, {\parallel a}}}$=1/4.
Therefore, the unit cell area reduction amounts to $S/S_0=0.25$ and the length $L/L_0=0.5$.
Similarly, we obtain for the $b$-axis 
$\alpha^b_0$=23T/K, $\kappa$=2.7K/$\mu_{\rm B}$ and $\chi_b$=0.013$\mu_{\rm B}$/T,
leading to $H^b_{\rm c2}/{H^{ \rm orb}_{\rm c2, {\parallel b}}}$=0.56.
Thus $S/S_0\sim0.56$ and the length $L/L_0\sim0.75$ for the $b$-axis.
These huge reductions are compared with the area reduction $\sim 15\%$
seen in the spin singlet superconductor TmNi$_2$B$_2$C due to the Pauli paramagnetic effect~\cite{Tm}
although the reduction mechanisms between them are completely different.

\section{Conclusion and summary}

We have discovered a novel mechanism to understand the upper critical field $H_{\rm c2}$ suppression
from its orbital limit in a spin triplet superconductor with the equal spin pairs and apply it to
the heavy Fermion superconductor UTe$_2$.
It is found that this $H_{\rm c2}$ suppression mechanism works well for UTe$_2$
and uncovers several mysteries associated with the anomalous $H_{\rm c2}$ behaviors
in UTe$_2$.
Notably, the remarkable $H_{\rm c2}$ enhancement observed in $H$$\parallel$$b$-axis 
is closely tied up with the present $H_{\rm c2}$ suppression. They occur in pair and
are different aspects with the same origin, namely the non-unitary state realized in UTe$_2$
is directly coupled with the underlying magnetization coming from the 5f localized moment.
The field induced moment controls $H_{\rm c2}$ in the system, either to suppress when 
the Cooper pair polarization is antiparallel or to enhance it when parallel.
In other words, it lets us monitor the Cooper spin orientation through $H_{\rm c2}$, 
providing a valuable monitoring tool other than the Knight shift experiment.

The identified non-unitary pairing symmetry is described by $(\hat{b}+i\hat{c})k_a$,
which is the so-called $\beta$ phase in the superfluid $^3$He~\cite{3he} and works quite successfully for various aspects of 
the observed phenomenology in UTe$_2$ in a consistent manner. This state breaks the time reversal symmetry
and the line node gap structure, which is classified group-theoretically ($^3$B$_{3u}$) in the assumption that 
the spin-orbit coupling is finite, not infinitely strong~\cite{annett}.

\section*{Acknowledgements}
The author sincerely thanks D. Aoki, K. Ishida, S. Kitagawa, S. Kittaka, Y. Tokunaga, A. Miyake, 
Y. Haga, H. Sakai, Y. Tokiwa, M. Kimata, and H. Matsumura for discussions of their experiments and 
Y. Tsutsumi for theoretical collaborations.
This work is supported by JSPS KAKENHI, No.~21K03455.

\end{document}